\documentclass[preprint,12pt]{elsarticle}

\usepackage{fullpage}
\usepackage{siunitx}
\usepackage{cancel}
\usepackage[usenames,dvipsnames]{xcolor}
\usepackage{multirow}

\usepackage{gensymb}

\usepackage{alltt}
\usepackage{upquote}
\usepackage{float}
\usepackage{array}
\usepackage{pgfplots}
\usepackage{pgfplots}
\pgfplotsset{compat=1.18}
\usetikzlibrary{positioning}
\usepackage{graphicx}
\usepackage[toc,page]{appendix}
\usepackage{amsmath,mathtools}
\usepackage{amsthm}
\usepackage{amsfonts}
\usepackage{enumitem}
\usepackage{caption}
\usepackage{booktabs} 
\usepackage{scalerel}
\usepackage{hyperref}
\usepackage{algorithm}
\usepackage{multicol}
\usepackage{mdframed}
\usepackage{mathtools}
\usepackage{enumerate}
\theoremstyle{remark}

\usepackage{natbib}
\usepackage{tabularx}

\usepackage{lineno}

\journal{Heat and Fluid Flow}

\begin{document}

\begin{frontmatter}



\title{An Efficient and Accurate Surrogate Modeling of Flapping Dynamics in Inverted Elastic Foils using Hypergraph Neural Networks}

\author[ubc]{Aarshana R. Parekh\corref{cor1}}
\ead{aarshana.parekh@.ubc.ca}
\cortext[cor1]{Corresponding author}
\author[ubc]{Rui Gao}
\author[ubc]{Rajeev K. Jaiman}
\ead{rjaiman@mech.ubc.ca}
\address[ubc]{Department of Mechanical Engineering, The University of British Columbia, Vancouver, BC V6T 1Z4}

\begin{abstract}
Cantilevered elastic foils or panels can undergo self-induced, large-amplitude flapping oscillations when submerged in a flowing fluid. This canonical phenomenon of fluid-structure interaction, commonly observed in nature, such as in the fluttering leaves or the movement of fish fins, offers a promising mechanism for marine energy harvesting. Through repetitive flapping, elastic foils harness kinetic energy from steady currents and wave-driven flows, with efficiency governed by oscillation amplitude and frequency. While high-fidelity simulations are essential for predicting and optimizing these parameters, they remain prohibitively expensive over wide parametric spaces.
To address this challenge, we develop a novel graph neural network (GNN) surrogate for the inverted foil problem, which is modeled as an elastically mounted rigid foil undergoing trailing-edge pitching in uniform flow. The coupled fluid-structure dynamics are resolved using a Petrov-Galerkin finite element method with an arbitrary Lagrangian-Eulerian formulation, providing high-fidelity data for training and validation. The surrogate employs a rotation-equivariant, quasi-monolithic GNN architecture: structural mesh motion is compressed via proper orthogonal decomposition and advanced through a multilayer perceptron, while the GNN evolves the flow field consistently with system states.
Specifically, this study extends the hypergraph framework to flexible, self-oscillating foils, capturing the nonlinear coupling between vortex dynamics and structural motion. The GNN surrogate achieves less than 1.5\% error in predicting tip displacement and force coefficients over thousands of time steps, while reproducing dominant vortex-shedding frequencies with high accuracy. The model captures the energy transfer metrics within 3\% of full-order simulations, demonstrating both accuracy and long-term stability. The results demonstrate a new and efficient surrogate for long-horizon prediction of unsteady flow-structure dynamics in energy harvesting systems.
\end{abstract}




\end{frontmatter}



\section{Introduction}\label{sec:introduction}

Flexible foils or panels submerged in a flowing fluid can undergo self-induced large-amplitude flapping oscillations. This canonical fluid-structure interaction phenomenon has been widely studied due to its prevalence in nature and diverse engineering applications, including increasing speed of paper printing \citep{Watanabe2002AFlutter}, nuclear reactor plate assemblies\citep{Guo2000StabilityFlow}, and flow control devices\citep{lucey1998excitation}. Largely, research has focused on suppressing these vibrations to reduce mechanical fatigue and extend the operational lifetime of structures. However, there are also applications where the phenomenon is advantageous, such as animal locomotion\citep{Zhang2000FlexibleWind}. In recent years, there has been a shift towards harnessing rather than mitigating these oscillations, with studies demonstrating their potential to improve heat transfer \citep{deng2022influence} and enhance the mixing process in channel flows \citep{ali2015heat}. In particular, self-sustained flapping of flexible foils in the inverted configuration, i.e., foils clamped at the trailing edge, has emerged as a promising mechanism for extracting fluid kinetic energy from steady currents and wave-driven flows\citep{allen2001energy,
Michelin2013EnergyFlows, Orrego2017HarvestingFlag,Alam2021EnergyFoil}. 

The inverted foils interact with the surrounding flow and undergo a large amplitude periodic flapping motion over a range of flow velocities. The oscillatory motion of these flexible foils can be harnessed to generate electric energy through piezoelectric composites\citep{Orrego2017HarvestingFlag}. Foil-based energy harvesting devices can be effective in air and water environments, while offering a lower environmental impact compared to traditional wind or marine turbines. Because the oscillations are self-sustained, meticulous calibration and active control strategies could enable continuous extraction of fluid kinetic energy through synchronized fluid-structure coupling. 
 Foil-based energy harvesters typically generate power on the micro- to milliwatt scale \citep{Shoele2016EnergyFlag, Orrego2017HarvestingFlag}. A piezoelectric energy harvester in which an elastic beam, instrumented with piezoelectric layers, undergoes large-amplitude flow-induced flapping oscillations. The continuous deformation of the beam under vortex-induced and flutter-type excitations generates strain in the piezoelectric patches, which is transduced into electrical energy. By embedding such piezoelectric mechanisms within flexible foils, the kinetic energy of ambient flows can be harvested at small scales to power devices that operate in this power range, such as remote sensors, data transmitters, buoys, and small-scale portable electronics \citep{Uihlein2016WaveTechnology}. These small-scale devices can be utilized in various applications, including marine environmental sensing and remote process monitoring. The power output of foil-based harvesters depends on the amplitude and frequency of oscillation, which in turn are governed by complex fluid–structure interactions. Accurate modeling of this coupling across a range of design configurations and flow conditions is therefore essential. Moreover, real-time prediction and control of flapping dynamics are critical to enhancing conversion efficiency and enabling practical deployment.

 State-of-the-art computational modeling approaches in computational fluid dynamics (CFD) and finite element analysis (FEA) are widely used for engineering analysis and design of coupled fluid-structure systems, such as the flexible foil. These high-fidelity tools, based on solutions of partial differential equations, capture the intricate dynamics of flapping motions and assess the hydrodynamic performance of different designs. While highly accurate, these approaches are computationally expensive, limiting their application to downstream tasks such as design optimization and control. Motivated by the need for faster and scalable modeling, we develop a data-driven surrogate model that accurately predicts the flapping dynamics of inverted foils at a fraction of the computational cost. Furthermore, data-driven simulation of fluid-structure interactions facilitates the development of small-scale intelligent devices such as sensors and buoys, enabling real-time, cost-effective, and autonomous monitoring of marine environments. These advancements tackle critical challenges in environmental sustainability, maritime security, and oceanographic research, fostering scalable solutions for efficient data acquisition and management. 

Data-driven models offer a promising pathway to overcome the computational limitations of traditional CFD and FEA in optimization and control tasks. Such models can be used in an offline-online manner. In the offline stage, the data-driven model is trained to learn a low-dimensional representation of the system states from the high-fidelity simulation data. Based on this training process, the model can now make efficient predictions in the online stage, thereby enabling iterative design optimization and real-time control for small-scale devices. This approach not only accelerates the improvement of foil-based energy harvesters but also makes their deployment in practical systems more feasible. Over the past decade, deep learning methods have been increasingly investigated as efficient alternatives to conventional CFD and FEA for fluid-structure interaction problems. Among the various deep learning approaches explored for fluid-structure systems, convolutional neural networks have received particular attention due to their ability to extract local flow features and reconstruct high-dimensional states efficiently.

Convolutional neural network (CNN) models have been applied to a range of fluid flow systems, including two-dimensional \citep{ thuerey2020deep, bukka2021assessment, pant2021deep} and three-dimensional flows over fixed bodies \citep{gupta2022three}, underwater noise propagation \citep{mallik2022predicting}, and fluid–structure interaction problems \citep{miyanawala2019hybrid, gupta2022hybrid}. In these approaches, autoencoders form nonlinear approximations of the coupled fluid–structure system. Combined with convolution operations that enforce locality and translational equivariance, encoder–propagator–decoder architectures efficiently compress and reconstruct high-dimensional flow states. Despite these advantages, CNNs face inherent challenges in resolving fluid-structure interfaces, particularly for thin bodies undergoing large deformations. Since convolution relies on a uniform Cartesian grid, the mesh resolution near the foil must match that of the far field. For thin foils, this requires a prohibitively dense grid to accurately resolve the surface. In contrast, coarser grids yield poor interface fidelity and unreliable estimates of critical quantities such as fluid loading. Hybrid approaches based on proper orthogonal decomposition (POD) \citep{miyanawala2019hybrid,reddy2019reduced}, body-fitted structured grids \citep{chen2023towards}, and interpolation or projection schemes \citep{gupta2022hybrid} partially alleviate these issues but remain constrained by their reliance on uniform grids, limiting their effectiveness for complex geometries.

To address the drawbacks of CNNs, this work leverages the recently introduced graph neural networks (GNNs) to model fluid-structure interaction dynamics. GNNs, introduced within the geometric deep learning framework, operate directly on graph representations that can be constructed from unstructured meshes. The mesh points of the computational grid correspond to the graph nodes, with the graph edges encoding connectivity. This representation enables locally refined resolution in critical regions, such as near-thin structures and moving boundaries, without relying on uniform Cartesian grids. GNNs, specifically the encode-decode-process architectures that employ generalized message passing and neighborhood aggregation, have shown strong capabilities in learning irregular spatio-temporal data. They have been increasingly applied to a wide range of fluid flow applications, including flow field super-resolution  \citep {belbute2020combining}, reacting flows \citep{xu2021conditionally}, flow around rigid bodies \cite{pfaff2020learning,lino2022multi} as well as fluid-acoustic shape optimization \citep{hadizadeh2024graph}. 

Recently, Gao \textit{et al.} \citep{gao2024finite} developed a finite element-inspired hypergraph neural network framework that uses a node–element hypergraph instead of a simple pairwise graph. Hypergraphs expand the expressivity of GNNs by replacing edges with hyperedges that connect multiple nodes simultaneously within a single interaction. This higher-order structure enables the model to capture multi-node physical couplings that standard GNNs cannot represent directly. Information is aggregated at the element level, and message passing mimics the local stiffness assembly process of finite element methods, thereby preserving geometric fidelity and local conservation while remaining rotation-equivariant. Recent studies have established the effectiveness of hypergraph neural networks for complex fluid flow problems \cite{gao2024finite,gao2024towards}, and fluid structure interaction systems\cite{gao2024predicting}. However, these works have only considered stationary geometries or bluff-body oscillators. The application of GNN/hypergraph-based surrogate models to flexible/self-oscillating foils, where the large deformation and wake structure coupling dominate the dynamics, has not been explored in a systematic manner. For instance, Pfaff \textit{et al.}\cite{pfaff2020learning} simulated the dynamics of a flexible flag in their work, but the GNN was limited to the flag itself, while the fluid flow surrounding the foil was not simulated. 

To the best of the authors’ knowledge, this work represents the first systematic investigation of the hypergraph neural networks to flexible inverted foils, a canonical configuration for flow-energy harvesting. The proposed surrogate captures the large-amplitude, self-sustained flapping of an elastically mounted rigid foil, where strongly coupled, nonlinear fluid–structure interactions govern the transfer of kinetic energy to the structure, marking a significant step toward data-driven modeling of flow-energy harvesters. Specifically, the coupled fluid–structure dynamics of the inverted foil are modeled using a recently developed quasi-monolithic GNN framework \citep{gao2024predicting}.
 The GNN architecture is based on the Arbitrary Lagrangian-Eulerian (ALE) formulation and consists of two sub-networks, one for the solid and one for the fluid, respectively. In the solid sub-network, the mesh and solid movements are first reduced to a lower dimension through a complex-valued proper orthogonal decomposition. The low-order POD coefficients are then propagated through time by a multilayered perceptron. The fluid sub-network comprises the finite element-inspired GNN ($\phi$-GNN), which updates the fluid state based on the state of the entire fluid-structure system \cite{gao2024finite}. In this architecture, the moving fluid mesh and solid states are combined, while the fluid states are updated separately, resulting in a quasi-monolithic framework. 

 Unlike CNN surrogates limited to structured domains, and message-passing GNNs prone to over-smoothing, the current $\phi$-GNN incorporates finite-element shape functions into hypergraph convolutions. This structure preserves local conservation and geometric fidelity, which is crucial for long-horizon prediction of inverted foil dynamics. This $\phi$-GNN  model is applied to a simplified setup of an elastically mounted rigid foil subject to uniform flow. The rigid foil undergoes a pitching motion about its trailing edge, i.e., an inverted foil configuration. Our analysis demonstrates that the GNN-based surrogate model efficiently predicts the flapping motion of the foil with strong agreement with ground truth data. This model delivers stable and accurate rollout predictions for at least thousands of steps. Although the present study focuses on a rigid pitching inverted foil, the underlying vortex-induced dynamics are directly relevant to piezoelectric and flexible foil harvesters, where accurate surrogate predictions of lift, displacement, and energy transfer metrics are essential.

This article is organized as follows. Section \ref{Sect:2} introduces the numerical frameworks employed in this study. First, we present the high-fidelity solver used to model the fluid-\\structure interaction dynamics of the inverted foil. We then describe the proposed GNN-based surrogate architecture for predicting the flapping dynamics. Section \ref{Sect:3} outlines the data generation and network training strategy, and Section \ref{Sect:4} reports the results that demonstrate the predictive capacity of the surrogate. Finally, Section \ref{Sect:5} summarizes the findings and highlights the directions for future research.
\section{Numerical Methodology} \label{Sect:2}
In this section, we first describe the high-fidelity solver used to generate ground-truth data. The framework couples the incompressible Navier–Stokes equations with a spring–\\
mass–damper system representing the structural dynamics. We then present the quasi-monolithic hypergraph neural network architecture employed to predict the flapping response of the inverted foil.

\subsection{High-Fidelity Model} \label{Subsect:2a}

We consider an elastically mounted rigid foil $
\mathrm{\Omega^{\mathrm{s}}}$ in the \textit{inverted} configuration that interacts with a uniform incompressible isothermal viscous fluid $\mathrm{\Omega^{\mathrm{f}}}$. Under the arbitrary Eulerian-Lagrangian formulation, the coupled fluid-structure system is modeled using a numerical scheme implementing Petrov-Galerkin finite elements and semi-discrete time stepping \cite{jaiman2016partitioned,jaiman2016stable}.

The viscous fluid flow is governed by the incompressible Navier-Stokes equations in the arbitrary Lagrangian-Eulerian reference frame over a time-dependent fluid domain $\Omega^{\mathrm{f}}(t)$ and expressed as follows:
\begin{align}
\rho^{\mathrm{f}}\left(\frac{\partial \mathbf{u}^{\mathrm{f}}}{\partial t}\Biggm|_{\hat{x}}+\left(\mathbf{u}^{\mathrm{f}} -\mathbf{w}\right) \cdot \nabla \mathbf{u}^{\mathrm{f}}\right) & =\boldsymbol{\nabla} \cdot \boldsymbol{\sigma}^{\mathrm{f}}+\mathbf{b}^{\mathrm{f}} \quad \text { on } \quad \Omega^\mathrm{f}(t), \\ \label{eq:fluid}
\nabla \cdot \mathbf{u}^{\mathrm{f}} & =0 \quad \text { on } \quad \Omega^{\mathrm{f}}(t), 
\end{align}
where $\mathbf{u}^{\mathrm{f}}(\mathbf{x},t)$ and $\mathbf{w}(\mathbf{x},t)$ are the fluid and mesh velocities at each spatial point $\mathbf{x}^{\mathrm{f}}\in \Omega^{\mathrm{f}}(t)$, respectively. 
$\mathbf{b}^{\mathrm{f}}$ is the body force per unit mass, and $\mathbf{\sigma}^{\mathrm{f}}$ is the Cauchy stress tensor for a Newtonian fluid, defined as:
\begin{align}
    \boldsymbol{\sigma}^{\mathrm{f}}=-p \mathbf{I}+\mu^{\mathrm{f}}\left(\nabla \mathbf{u}^{\mathrm{f}}+\left(\nabla \mathbf{u}^{\mathrm{f}}\right)^T\right), 
    \label{eq:Cauchy}
\end{align}
where $p$, $\mu^{\mathrm{f}}$, and $\mathbf{I}$ are the hydrodynamic pressure, the dynamic viscosity, and the identity tensor, respectively. The first term in eq.\ref{eq:fluid} corresponds to the partial time derivative of $\mathbf{u}^{\mathrm{f}}$ while the ALE referential coordinate, $\mathrm{\hat{x}}$, is kept fixed. 

An elastically mounted rigid foil immersed in a fluid flow is subjected to hydrodynamic loading that induces structural vibrations. The resulting pitching motion of the structure about its center of rotation is described by the following equation:
\begin{align}
    I_\theta\frac{d^2\theta^s}{dt^2}+ C_{\theta}\frac{d\theta^s }{dt}  + K_{\theta}(\theta^s-\theta_0) =\mathbf{M}^{\mathrm{f}} + \mathbf{M}^{b} \quad \text{on} \quad \Omega^{\mathrm{s}}, 
    \label{eq:solid}
\end{align}
where $I_\theta$, $C_\theta$, and $K_{\theta}$ denote the moment of inertia, damping, and spring stiffness vectors of the rotational degrees of freedom, respectively. $\theta^{\mathrm{s}}$  represents the angular displacement of the rigid foil at time $t$ whereas $\theta_{0}$ denotes its initial angular displacement. $\mathbf{M}^{\mathrm{f}}$ and $\mathbf{M}^{b}$ denote the torque acting on the foil due fluid traction  $(\mathbf{F}^{\mathrm{f}})$ and body forces $(\mathbf{b}^{\mathrm{s}})$ on the structure. 
Notably, the fluid-induced moment $\mathbf{M}^{\mathrm{f}}$ directly couples the solid equation with the fluid equations through the traction exerted on the fluid-structure interface.

The fluid and structural dynamics are coupled by enforcing continuity of velocity and traction along the fluid-structure interface. Let $\Gamma^{\mathrm{fs}} = \partial\Omega^f(0)\cap \partial\Omega^s(0)$ denote the interface at time $t=0$, and $\Gamma^{\mathrm{fs}} = \varphi^{\mathrm{s}}(\Gamma^{\mathrm{fs}},t)$ its position at time $t$. The coupled system must satisfy
\begin{align}
&\mathbf{u}^{\mathrm{f}}(\varphi^{\mathrm{s}}(z_0,t),t)  =\mathbf{u}^{\mathrm{s}}(z_0,t)  \\
&\int_{\varphi^{\mathrm{s}}(\gamma,t)} \boldsymbol{\sigma}^{\mathrm{f}}(\mathbf{x},t) \cdot \mathbf{n}\mathrm{d}\Gamma(\mathbf{x}) \; + \int_{\Gamma}F^{\mathrm{s}}\mathrm{d}\Gamma = 0 , \label{eq:interface}
\end{align}
where $\varphi^{\mathrm{s}}$ denotes the position vector mapping the initial position $\mathbf{z}_0$ of the structure to its position at time $t$, $\gamma$ is any part of the fluid-structure interface $\Gamma_{\mathrm{fs}}$ in the reference configuration. $\mathbf{u}^{\mathrm{s}}$ is the velocity of the solid body at time $t$ and $\mathbf{n}$ is the unit normal at the fluid-structure interface $\Gamma^{\mathrm{fs}} (t)$. $\mathrm{d}\Gamma$ denotes a differential surface area and $\varphi^{\mathrm{s}}(\gamma,t)$, its corresponding fluid part at time $t$. In eq. \ref{eq:interface}, the first term represents the traction forces exerted by the fluid on $\varphi^{\mathrm{s}}(\gamma,t)$, while the second term represents the total force exerted by the rigid body on $\varphi^{\mathrm{s}}(\gamma,t)$.

The motion of each spatial point within the fluid domain is explicitly updated to maintain kinematic consistency as the fluid-structure interface evolves over time. The weak form of Eq. \ref{eq:fluid} is discretized in space using $\mathbb{P}_2 / \mathbb{P}_1$ iso-parametric finite elements for velocity and pressure, and in time using a second-order backward differentiation scheme. A partitioned staggered approach is employed to couple the fluid and structural solvers. The robustness and accuracy of this numerical framework have been extensively validated in prior studies on diverse FSI problems \citep{Liu2014AEffects,jaiman2016stable}. For detailed algorithmic and implementation aspects, the reader is referred to Jaiman \textit{et al.}\citep{jaiman2016stable}. Importantly, within this ALE framework, the continuity of velocity and traction at the interface is enforced naturally through the finite element formulation, ensuring stable and accurate coupling between the fluid and structural fields.

\subsection{Data-Driven Surrogate Modeling}\label{Subsect:2B}
The coupled differential equations (\ref{eq:fluid}) and (\ref{eq:solid}) of the fluid-structure interaction problem can be expressed in abstract dynamical form as
 \begin{equation}
     \frac{d\mathrm{y}}{dt} = \tilde{\mathbf{F}}(\mathrm{y}), \label{eq:dynamical}
 \end{equation}
where $\mathbf{y}$ is the column vector representing the system's state with unknown degrees of freedom, and $\tilde{\mathbf{F}}$ is the vector-valued function describing the spatially discretized governing equations (\ref{eq:fluid}-\ref{eq:solid}). In the FSI system, the state vector comprises the fluid velocity $\mathbf{u}^{\mathrm{f}} =\left(\mathrm{u^{\mathrm{f}}_x},\;\mathrm{u^{\mathrm{f}}_y}\right)$, hydrodynamic pressure $p$, and the solid state. 
This dynamical system representation provides the foundation for developing a reduced-order surrogate model, wherein the GNN architecture is trained to approximate the nonlinear operator $\tilde{\mathbf{F}}$ and predict the long-horizon evolution of the coupled FSI states.

Given a constant time step $\delta t$ used for temporal discretization of the FSI system,  Eq. (\ref{eq:dynamical}) can be rewritten using the forward Euler integration scheme as:
\begin{align}
    \mathbf{y}\left(t_{n+1}\right)-\mathbf{y}\left(t_n\right)=\delta \mathbf{y}=\mathbf{F}\left(\mathbf{y}\left(t_n\right)\right), \label{eq:FEuler}
\end{align}
where $\mathbf{y}\left(t_n\right)$ denotes the system state at time step $t_n$, and $\mathbf{F}$ represents the full-order system state update function in the discretized time domain. In surrogate modeling, the objective is to approximate $\mathbf{F}$ with a learned function
\begin{align}
\mathbf{F}\left(\mathbf{y}\left(t_n\right)\right)=\mathbf{y}\left(t_{n+1}\right)-\mathbf{y}\left(t_n\right) \approx \delta \hat{\mathbf{y}}=\hat{\mathbf{F}}\left(\mathbf{y}\left(t_n\right), \theta\right),
\end{align}
where $\theta$ represents the parameters of the surrogate model $\hat{\mathbf{F}}$. These parameters are calculated and tuned to minimize the error between the approximation and the full-order model.
\begin{align}
    \theta=\arg _\theta \min (\mathcal{L}(\mathbf{F}, \hat{\mathbf{F}}(\theta))).\label{eq:surrogate}
\end{align}
where $\mathcal{L}$ represents the loss function. In this work, we adopt a hypergraph-based neural network as a surrogate $\hat{\mathbf{F}}$ to predict the flapping dynamics of the inverted foil in uniform flow. The computational mesh used for numerical simulations is mapped into a hypergraph representation, where nodes correspond to state variables and hyperedges encode geometric relationships. The hypergraph neural network leverages these mesh-based relationships to learn the surrogate operator and predict the temporal evolution of coupled fluid-structure states at all mesh nodes, based on information from previous time steps. The following subsections detail the adapted data-driven framework.
\subsection{Finite Element Inspired Hypergraph Neural Networks }\label{Secsec:2C}
Graph neural networks work on graphs $G=(\mathcal{V,E})$ where $\mathcal{V}$ is a set of nodes  $N = |\mathcal{V}|$, and  $\mathcal{E}$ is a set of edges $E= |\mathcal{E}|$, wherein a pair of nodes are connected through either a directed or undirected edge. A hypergraph $H = (\mathcal{V}, \mathcal{E})$ generalizes this structure by allowing each hyperedge $e \in \mathcal{E}$ to connect with more than two nodes. Unlike standard graphs, where edges capture pairwise relationships, hyperedges naturally encode higher-order interactions among multiple nodes, enabling hypergraph neural networks to represent richer geometric and topological relationships in the data.

We employ the recently developed finite-element-inspired hypergraph neural network architecture ($\phi$-GNN) \cite{gao2024finite} for modeling fluid-structure interaction systems. Consider a bounded two-dimensional spatial domain discretized into a finite element mesh, as illustrated in Fig. \ref{fig:FOMtoGNN}a. In the finite element method, mesh vertices represent nodes, and the cells formed by connecting these vertices define elements. In the $\phi$-GNN framework,  this finite element mesh is reformulated as a node–element undirected hypergraph (Fig. \ref{fig:FOMtoGNN}b). Each mesh element is converted into a hyperedge $e_s$ connecting all of its associated vertices, while each vertex becomes a hypergraph node. Additionally, explicit undirected element–node edges connect each element to its corresponding nodes.
The complete hypergraph is thus defined as $\mathcal{G} = (\mathcal{V}, \mathcal{E}_s, \mathcal{E}_v)$, where $\mathcal{V}$ is the set of nodes, $\mathcal{E}_s$ the set of element hyperedges, and $\mathcal{E}_v$ the set of element–node edges. This construction preserves the mesh topology and connectivity, while embedding the system states $\mathbf{y}$ into node, element, and element–node feature vectors. These feature vectors form the input representation for the $\phi$-GNN. The transformation of high fidelity of data into feature vectors is discussed further in Sec \ref{Subsect:2D}.
By leveraging this FEM-to-hypergraph mapping, the $\phi$-GNN inherits local conservation properties and naturally captures higher-order relationships among states, which are essential for accurate long-horizon prediction of fluid–structure dynamics.

The $\phi$-GNN architecture consists of three stages: encoding, processing, and decoding. In the encoding stage, the node–element hypergraph and its associated feature vectors are embedded into the network through encoder functions. Let $\mathit{v}_i$ denote the feature of node $i$, $\mathit{e}s$ the feature of element $s$, and $\mathit{e}{s,i}$ the feature of the element–node edge connecting node $i$ and element $s$. The encoding step is expressed as
\begin{align}
    \mathit{v}_i & \leftarrow g^{\mathit{v}}\left(\mathit{v}_i\right), \\
    e_{s} & \leftarrow g^{\mathit{e}}\left(e_{\mathit{s}}\right), \\
    e_{s, i} & \leftarrow g^{\mathit{e v}}\left(\mathit{e}_{\mathit{s}, i}\right),
\end{align}
where $g^{\mathit{v}}$, $g^{\mathit{e}}$, and $g^{\mathit{ev}}$ denote the encoder functions for the node, element, and element–node features, respectively. The update symbol $\leftarrow$ indicates that the feature on the left-hand side is replaced by the transformed representation on the right-hand side. These encoded features form the input to the processing stage, where message passing over the hypergraph captures higher-order interactions among nodes and elements.

In the processing stage, the encoded feature vectors are propagated through a sequence of generalized message-passing layers, each comprising an element update followed by a node update. In the $\phi$-GNN framework, the hypergraph message passing process mimics the local stiffness assembly process of the finite element method. For a detailed derivation of the message passing layers, the readers are encouraged to refer to Gao \textit{et al.} \cite{gao2024finite}.
For an element $s$ connecting four nodes $i,j,k,\text{and }l$, the element update stage and subsequent node update stage for each node $i$ can be written as
\begin{align}
    \mathit{e_s} & \leftarrow \operatorname{AGG}^{e}_{r} \left(\phi^{\mathit{e}}\left(\mathit{v}_{r},\mathit{e_s},\mathit{e_{s,r}}\right)\right), \\
    \mathit{v_i} & \leftarrow \operatorname{AGG}^{v}_{s} \left(\phi^{\mathit{v}}\left(\mathit{v}_{i},\mathit{e'_{s_{i}}},\mathit{e_{s_{i},i}}\right)\right).
\end{align}
Here, $\operatorname{AGG}$ denotes an aggregation function which is chosen to be the mean function in this work. $r=i,j,k,l$ denotes the node index, $\mathit{s_{i}}$ denotes any element connecting node $i$ with other nodes, and $\mathit{e}'_{\mathit{s_{i}}}$ represents the element feature updated by the previous element update stage. The element update $\phi^{\mathit{e}}$ and node update $\phi^{\mathit{v}}$ functions are node shared for different message passing layers.
The outputs at the end of message-passing layers are decoded to produce the final outputs of the network in the decoding stage. In this work, only outputs from the element note edges are needed; therefore, only the element-node decoder $\mathit{h^{ev}}$ is required
\begin{equation}
    \mathit{e_{s,i}} \leftarrow \mathit{h^{ev}}(\mathit{e_{s,i}}).
\end{equation}
The element-node outputs are further post-processed to transform into node-level output vectors. 

The encoder, decoder, element, and node update functions are implemented using multi-layered perceptions. A multi-layered perception (MLP), also known as a feedforward network, is one of the simplest forms of modern neural networks. An MLP with multiple layers is a network that maps the input vector to the output through linear transformations and non-linear activation functions. For an input vector $\boldsymbol{z}$, a multi-layered perception  $n_l-1$ hidden layers can be written as
\begin{equation}
    f(\boldsymbol{z})=f_{n_l} \circ f_{n_l-1} \circ f_{n_l-2} \circ \cdots \circ f_2 \circ f_1(\boldsymbol{z})
\end{equation}
in which $\circ$ denotes function composition. Each layer $f_i$ is defined as
\begin{align}
    f_i(\boldsymbol{z})= \begin{cases}\sigma_i\left(\boldsymbol{W}_i \boldsymbol{z}+\boldsymbol{b}_i\right) & i=1,2, \ldots, n_l-1 \\ \boldsymbol{W}_i \boldsymbol{z}+\boldsymbol{b}_i & i=n_l\end{cases}
\end{align}
with $\boldsymbol{W}_i$ and $\boldsymbol{b}_i$ denoting the trainable weight matrix and bias vector, respectively. The layer width is determined by the number of rows in $\boldsymbol{W}_i$, and $\sigma_i(\cdot)$ is the nonlinear activation function.  This composition of MLPs involving layers of transformations allows the network to extract hierarchical representations from the data. In the next section, we describe how the hypergraph framework is adapted for fluid-structure interaction problems.

\subsection{Hypergraph based Fluid-Structure Interaction Framework} \label{Subsect:2D}

 In this work, we model the fluid–structure interaction system of a flexible inverted foil undergoing flow-induced vibrations in a two-dimensional domain discretized with a body-fitted mesh using the ALE-based finite element formulation. At each mesh vertex, the system states, vertex coordinates, and the structure-to-fluid mass ratio $m^*$ are transformed into input features for the hypergraph neural network. The objective of the framework is to forecast the flow variables ${u_x, u_y, p}$ and mesh displacements ${\Delta x, \Delta y}$ at all vertices at time $t_{n+1}$, conditioned on the system states at the previous timestep. To achieve this, we adopt a quasi-monolithic formulation comprising two sub-networks: one capturing the temporal evolution of the mesh state and the other predicting the flow field. For clarity, the details of each sub-network are presented in the following subsections. While certain components build upon our earlier work \citep{gao2024predicting}, they are included here for completeness and continuity.

\subsubsection{Solid Sub-network}\label{SubSubsect:solidsub}

A hybrid approach, utilizing complex-valued proper orthogonal decomposition and multilayer perceptrons, evolves the solid/mesh states over time. Given a 2D mesh, the displacements $(\delta \mathbf{x}, \delta \mathbf{y})$ of vertices at each time step $t_n$ are assembled in the complex plane
\begin{align}
    \delta \mathbf{X}_{t_n}=\delta \mathbf{x}+\mathrm{i} \delta \mathbf{y}
\end{align}
The complex vectors of the training dataset are concatenated horizontally to construct vector $\mathbf{\delta X}$. The displacements of the mesh vertices are computed with respect to their initial positions, thereby ensuring that the mesh state predictions are translational invariant. Singular value decomposition is performed on $\mathbf{\delta X} \in \mathbb{C}^{P\times N}$ which results:
 \begin{equation}
     \mathbf{\delta X=U\mathrm{\Sigma} V}^{H}, 
 \end{equation}
where $P$ represents the number of mesh vertices and $N$ represents the number of time steps considered for decomposition. $H$ denotes the conjugate transpose operation.
$\mathbf{U}$ denotes the left unitary matrix with each column representing modes, with the leftmost columns capturing the most variance in the data. The singular values are located on the diagonal of matrix $\mathrm{\Sigma}$, arranged in descending order. 

Assuming that the solid motion is low-rank, a reduced-order approximation of the ALE mesh displacement field can be constructed by projecting the data onto the first $n_r$ modes. To preserve rotational equivariance in the predictions, we adopt a complex-valued POD formulation instead of the standard real-valued variant \citep{gao2024predicting}. The projection–truncation operation is expressed as
\begin{align}
    \mathbf{C}^{(1: n_r)}=\left(\mathbf{U}^{(1: n_r)}\right)^H \delta \mathbf{X}, \\
\delta \hat{\mathbf{X}}=\mathbf{U}^{(1:n_r)} \mathbf{C}^{(1:n_r)},
\end{align}
where $\mathbf{U}^{(1:n_r)}$ denotes the truncated basis formed by the first $n_r$ columns of $\mathbf{U}$, $\mathbf{C}^{(1:n_r)}$ contains the modal coefficients, and $\delta \hat{\mathbf{X}}$ represents the low-order reconstruction of $\delta \mathbf{X}$. These reduced coefficients $\mathbf{C}^{(1:n_r)}$ are propagated forward in time using an MLP, providing a compact and efficient representation of mesh dynamics within the GNN framework. The temporal evolution of the reduced system is approximated as
\begin{align}
    \delta \mathbf{X}_{t_n} \approx \mathbf{U}^{\left(1: n_r\right)} \mathbf{C}_{t_n}^{\left(1: n_r\right)}, 
\end{align}
where $\mathbf{C}_{t_n}^{\left(1: n_r\right)}$ is the truncated POD coefficient matrix at each timestep  $t_n$ defined as:
\begin{align}
    \mathbf{C}_{t_n}^{\left(1: n_r\right)}=\boldsymbol{\rho}_{t_n}^{\left(1: n_r\right)} \exp \left(\mathrm{i}\boldsymbol{\eta}_{t_n}^{\left(1: n_r\right)}\right).
\end{align}
with $\boldsymbol{\rho}_{t_n}^{(1:n_r)}$ and $\boldsymbol{\eta}_{t_n}^{(1:n_r)}$ denoting the amplitude and phase components, respectively.

Multilayered perceptrons are employed to predict the reduced coefficients $\mathbf{C}_{t_{n+1}}^{\left(1: n_r\right)}$ at time step $t_{n+1}$. Information from multiple history time steps $h$ is provided as input to the network to make these predictions. Given information of coefficients within the history window $h$, the first changes in moduli $\delta\boldsymbol{\rho}_{t_n}$  and arguments $\delta\boldsymbol{\eta}_{t_n}$ are computed for consecutive time steps within this window. 
 \begin{align}
     & \delta \mathbf{\rho}_{t_n}^{(1: n_r)}=\mathbf{\rho}_{t_n}^{(1: n_r)}-\mathbf{\rho}_{t_{n-1}}^{(1: n_r)}, \\
& \delta \mathbf{\eta}_{t_n}^{(1: n_r)}=\mathbf{\eta}_{t_n}^{(1: n_r)}-\mathbf{\eta}_{t_{n-1}}^{(1: n_r)}.
 \end{align}
The computed increment in moduli $\bigg[\delta \mathbf{\rho}_{t_n}^{(1: n_r)}, \delta \mathbf{\rho}_{t_{n-1}}^{(1: n_r)}, \dots, \delta \mathbf{\rho}_{t_{n-h+1}}^{(1: n_r)}\bigg]^T$ and arguments \\$\bigg[\delta \mathbf{\eta}_{t_n}^{(1: n_r)}, \delta \mathbf{\eta}_{t_{n-1}}^{(1: n_r)}, \dots, \delta \mathbf{\eta}_{t_{n-h+1}}^{(1: n_r)}\bigg]^T$ are provided as inputs to the MLPs to predict the increments in $\hat{\mathbf{\rho}}^{(1:n_r)}_{t_{n+1}}$ and $ \hat{\mathbf{\eta}}^{(1:n_r)}_{t_{n+1}}$
following the forward Euler scheme Eq.(\ref{eq:FEuler}):
\begin{align}
\left[\begin{array}{c}
\hat{\mathbf{\rho}}_{t_{n+1}}^{\left(1: n_r\right)} \\
\hat{\mathbf{\eta}}_{t_{n+1}}^{\left(1: n_r\right)}
\end{array}\right]=\left[\begin{array}{c}
\hat{\mathbf{\rho}}_{t_n}^{\left(1: n_r\right)} \\
\hat{\mathbf{\eta}}_{t_n}^{\left(1: n_r\right)}
\end{array}\right]+\left[\begin{array}{c}
\delta \hat{\mathbf{\rho}}_{t_{n+1}}^{\left(1: n_r\right)} \\
\delta \hat{\mathbf{\eta}}_{t_{n+1}}^{\left(1: n_r\right)}
\end{array}\right] .
\end{align}
The predicted low-order coefficients are multiplied by the corresponding modes to recover the predicted mesh displacement vectors at each timestep as follows:
\begin{equation}
    \delta \hat{\mathbf{X}}_{t_n}=\mathbf{U}^{\left(1: n_r\right)} \hat{\mathbf{\rho}}_{t_n}^{\left(1: n_r\right)} \exp \left(\mathrm{i} \hat{\mathbf{\eta}}_{t_n}^{\left(1: n_r\right)}\right),
\end{equation}
with the real and imaginary parts of these vectors corresponding to the $\mathbf{x}$ and $\mathbf{y}$ components of the mesh displacement, respectively. The employed autoregressive approach provides a simple and efficient framework by reusing a single-step prediction model without the need for complex sequence encoders or decoders. This makes it well-suited for real-time prediction and control, as outputs are generated incrementally without requiring the full temporal history. Moreover, the use of complex-valued POD naturally enforces rotational symmetry in flow-induced vibrations. By separating phase and amplitude, the model compactly encodes cyclic oscillations while avoiding the higher-dimensional latent space required in a purely real-valued formulation.

\subsubsection{Fluid sub-network} \label{SubSubsect:fluidsub}
\begin{figure*}[ht]
    \centering
    \includegraphics[width=\textwidth]{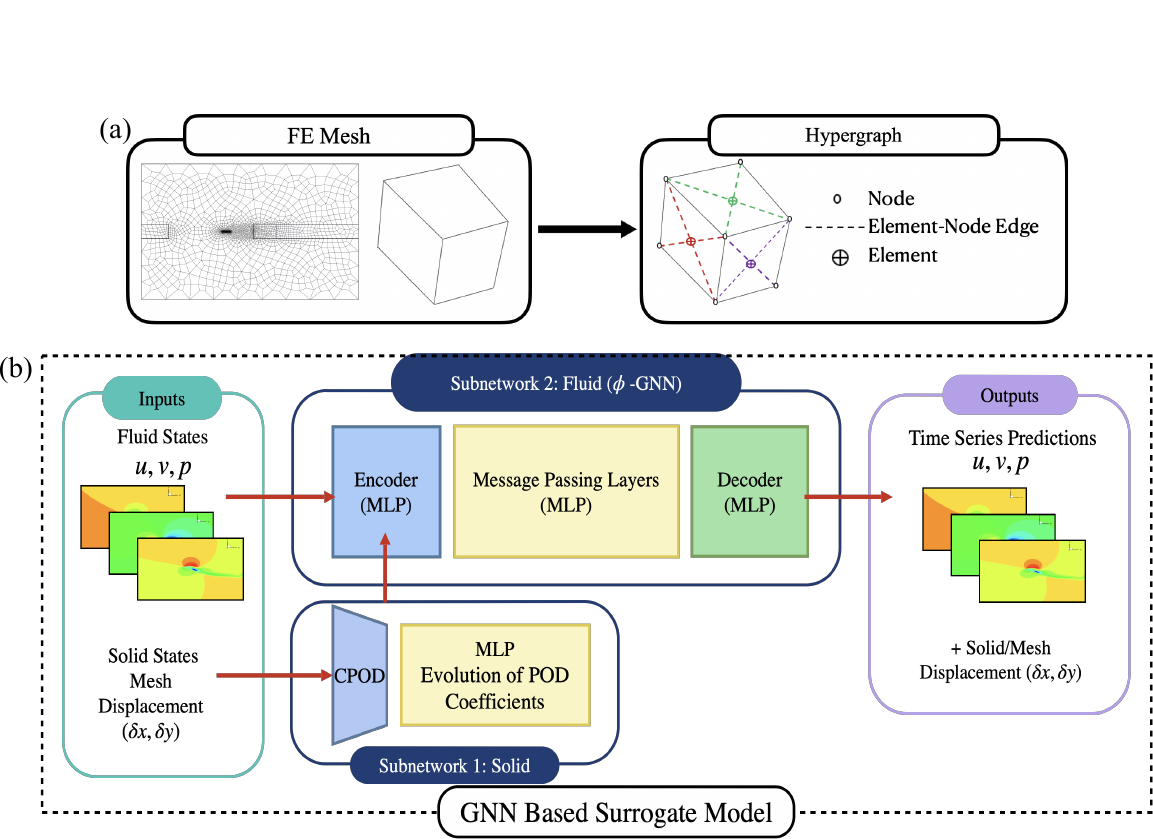} 
    \caption{(a) Conversion from computational mesh to node-element hypergraph, (b) A diagram illustrating the graph neural network-based reduced order model using the quasi-monolithic framework trained on full order model time-series data}
    \label{fig:GNN}
\end{figure*}

The hypergraph is constructed from the computational mesh, as shown in Fig. \ref{fig:GNN}a. It follows the encode–process–decode architecture illustrated in Fig. \ref{fig:GNN}b, similar to other networks discussed in Sec. \ref{sec:introduction}. MLPs are employed as encoders, decoders, and update functions within the message-passing layers of this sub-network. The predicted flow features at each timestep are post-processed from the network output. Next, we briefly describe the features of the GNN constructed from the high-fidelity data. The initial mesh is used to define the geometry features of the GNN. Given an element $s$, the coordinates of its center $(x_s, y_s)$ are calculated as the mean of the connecting node coordinates $(x_r, y_r)$, where $r$ indexes the element’s nodes. Local node coordinates are then defined relative to the element center as $(x_{s,r}, y_{s,r})$. The local coordinate length $L_{s,r}$, corner angle $\theta_{s,r}$, and element area $A_{s,r}$ together form the geometry features of the mesh. In addition to mesh shape, boundary conditions are encoded as one-hot vectors $\gamma_i$ for each node $i$.

The ALE mesh displacements $(\delta \mathrm{x}, \delta \mathrm{y})$ and flow velocities $( \mathrm{u_{x}}, \mathrm{u_{y}})$  are  projected to the local co-ordinate directions $\mathbf{x_{s,i}}$ for each node $i$ to form scalar features $d_{s,i}$ and $f_{s,i}$, respectively.
\begin{align}
    d_{s,i} = \frac{1}{\mathrm{L}_{s,i}}
\begin{bmatrix}
\mathrm{\delta x} \; \mathrm{\delta y}
\end{bmatrix}
\cdot
\begin{bmatrix}
\mathrm{x_{s,i}} \\ \mathrm{y_{s,i}}
\end{bmatrix}, \quad
u_{s,i} = \frac{1}{\mathrm{L}_{s,i}}
\begin{bmatrix}
\mathrm{u_x} \; \mathrm{u_y}
\end{bmatrix}
\cdot
\begin{bmatrix}
\mathrm{x}_{s,i} \\ \mathrm{y}_{s,i}
\end{bmatrix},
\end{align}
The scalar pressure $p_i$ at each node and the structure-to-fluid mass ratio $(m^*)$ for each simulation case are also included as features of the GNN.

The formulated features are attached to the hypergraph as follows. The boundary condition feature vector $\gamma_{i}$ is attached as a node feature $v_{i}$, and $A_{s}$ is naturally considered as the element feature. Note that instead of directly including  $A_{s}$ and  $\mathrm{L}_{s,i}$, their natural logarithm is considered to account for large variations in values of area and length across the mesh. The pressure value $p_i$ is gathered on each element node edge $p_{s,i}$ and non-dimensionalized using $\rho^{\mathrm{f}}$ and freestream velocity $\mathrm{U_\infty}$, following which its signed square root is concatenated with other available information to form the element-node edge feature vector as follows:
\begin{equation}
e_{s, i}=\left[
u_{s, i}\quad \operatorname{sgn}\left(p_{s, i}\right) \sqrt{\frac{2\left|p_{s, i}\right|}{\rho^f \mathrm{U}_{\infty}^2}}\quad   d_{s, i}\quad   m^{*}\quad   \ln \mathrm{L}_{s, i}
 \quad  \cos \theta_{s, i}\right]^T,
\end{equation}
where sgn$(\cdot)$ denotes the sign function and $m^*$ represents the structure-to-fluid mass ratio for the system.
The $\phi$-GNN $(\hat{\mathbf{G}})$ iteratively predicts the increment in feature vectors using the forward Euler scheme Eq. (\ref{eq:FEuler}) at each time step as:
\begin{align}
    e_{s,i}^{t+1} \approx e_{s,i}^t + \hat{\mathbf{G}}(v_i, e_{s,i}, e_s),
\end{align}

The hypergraph then updates the projected velocity and pressure values, encoded as features, at the element–node edges at each timestep. The updated element-node features are post-processed to retrieve the velocity and pressure values at each node. The scalar pressure at given node $i$ is computed by taking the mean across all the element nodes connecting to that node. The velocity and displacement are reverse-projected using the Moore-Penrose pseudo-inverse. A schematic of the quasi-monolithic architecture is shown in Fig. \ref{fig:GNN}(b) with red arrows representing data flow during temporal rollout predictions. In the following section, we provide the results obtained while using the hypergraph neural network to predict the flapping dynamics of inverted foils.
\section{Data Generation and Training} \label{Sect:3}
This section first describes the ground-truth data generation process and the organization of the training and testing datasets. We then outline the model architecture along with the training and testing strategy. We begin with the generation of the high-fidelity fluid–structure interaction dataset.

\subsection{Ground Truth Data Generation}\label{Subsect:3a}
\begin{figure}[ht]
\centering\includegraphics[width=0.65\linewidth]{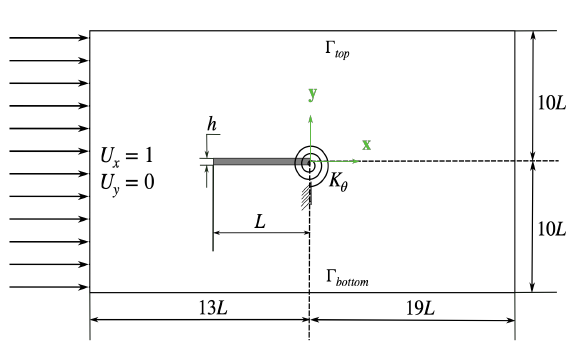}
\caption{Schematic of the computational domain used for generating the ground truth data}\label{fig:schem}
\end{figure}
\begin{figure}[ht]
\centering\includegraphics[width=0.45\linewidth]{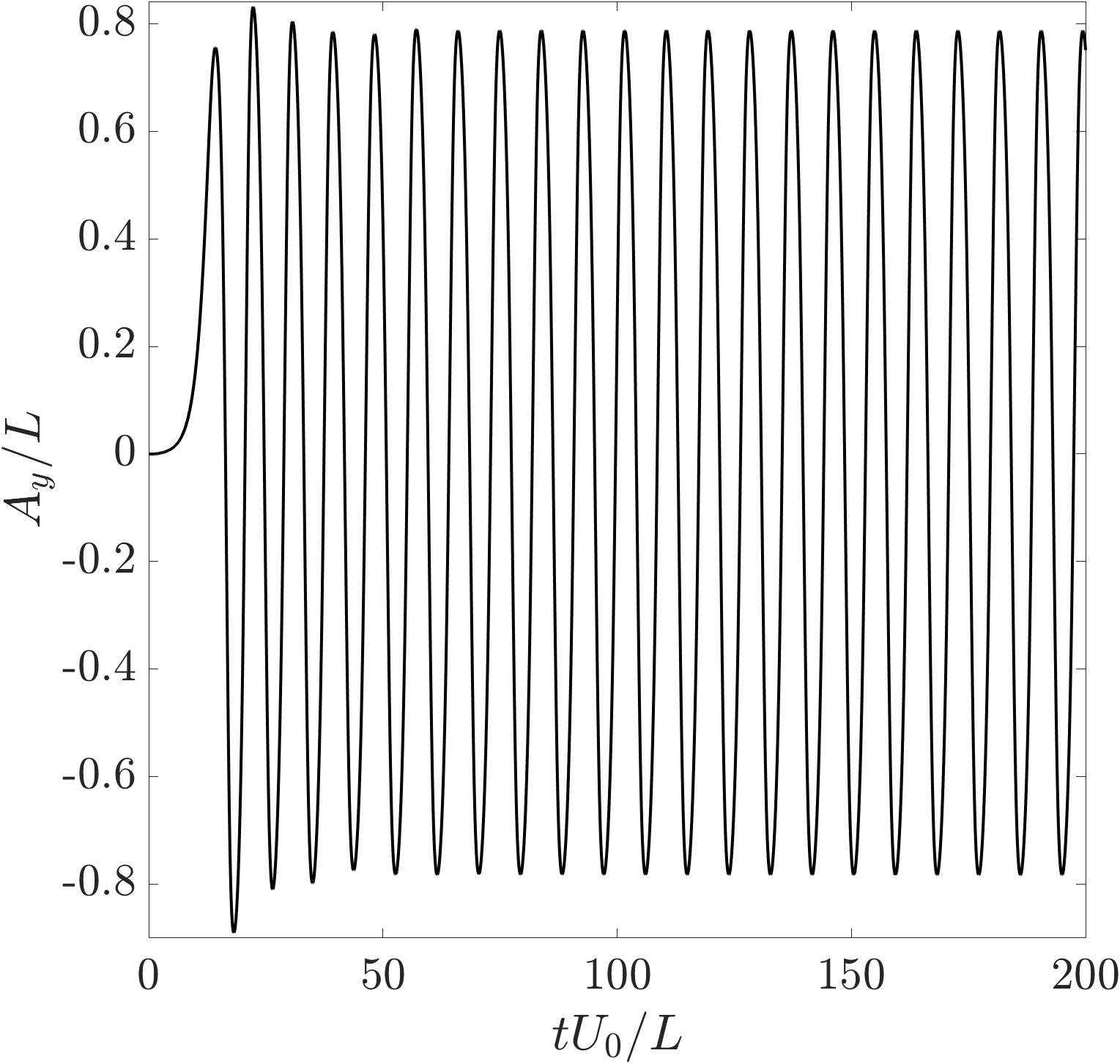}
\caption{Transverse tip displacement of the rigid foil at $m^* = 1$, $Re = 1000$ and $K^*_\theta = 2.4$  }\label{fig:osci}
\end{figure}
The ground truth datasets for training and testing the model are generated using high-fidelity numerical simulations with the body-fitted ALE formulation, as described in Sect. \ref{Subsect:2a}.
To evaluate the model's ability to predict the flapping dynamics, we consider the case of a flow around an elastically mounted rigid foil of length $L$ and height $h = 0.01L$ as shown in the schematic in Fig. \ref{fig:schem}. The trailing edge of the rigid plate is located at the origin of the Cartesian coordinate system. The distances between the upstream and downstream boundaries are $13L$ and $19L$ from the trailing edge, respectively. The distance between the side walls is $20L$. The foil is free to undergo pitching motion about its trailing edge. The inlet velocity is set to unity, with a no-slip boundary condition imposed on the plate surface, while the top and bottom boundaries are defined as slip walls.
In this setup, similar to the flexible inverted flag \cite{Kim2013FlappingFlag, Gurugubelli2015Self-inducedFlow, leontini2022dynamics},  the dynamics of the rigid foil are driven by three non-dimensional parameters, namely the Reynolds number $(Re)$, the structure to fluid mass ratio $(m^*)$, and non-dimensional spring stiffness $(\mathrm{K}^{*}_{\theta})$, defined as
\begin{align}
    Re = \frac{\rho^{\mathrm{f}}\mathrm{U}_{\infty}L}{\mu^{\mathrm{f}}},\quad m^* = \frac{\rho^{\mathrm{s}}h}{\rho^{\mathrm{f}}L}, \quad K^{*}_{\theta} = \frac{K_{\theta}}{\frac{1}{2}\rho^{\mathrm{f}}\mathrm{U}^{2}_{\infty}L^2},
\end{align}
where $\rho^{\mathrm{s}}$ is the solid density. 

To determine the adequacy of the computational mesh, a convergence study is performed using three different mesh resolutions: $M1$, $M2$, and $M3$.
An elastically mounted isolated inverted rigid foil is considered at $Re = 1000$, $m^{*} = 1$, and $K^{*}_{\theta}=2.4$. Table \ref{tab:Tab1mesh} lists the root mean squared (r.m.s) values of the normalized transverse tip displacement ($A^{rms}_{y}/L$), lift coefficient ($C^{rms}_{l}$), the mean drag coefficient ($\overline{C_{d}}$), non-dimensional flapping frequency $f^{*}_{A_y} = f_{A_y}L/U$ as well as the number of nodes on the plane of the mesh along the z-axis for the different meshes.  The bracketed values denote the percentage difference in numerical simulations compared to mesh $M3$. The relative errors with mesh $M2$ are less than $2\%$; therefore, all subsequent simulations are carried out using mesh $M2$. It is important to note that only a single mesh-element layer is used in the spanwise direction for all three meshes. Thus, while the computational domain is three-dimensional, the dynamics captured in this study are essentially two-dimensional. This validated mesh configuration provides the ground-truth data for training the surrogate model with confidence in numerical accuracy.
\begin{table*}[ht]
\centering
\begin{tabular}{lccc}
          & $M1$ &$M2$ & $M3$ \\ \hline
Nodes              & 23340            & 46075            & 93260 \\
$A^{\mathrm{rms}}_{y}/L$ & 0.8385 (3.6765\%) & 0.8325 (0.7208\%) & 0.8385 \\
$\overline{C_{d}}$ & 1.5333 (7.7634\%) & 1.4018 (1.4737\%) & 1.4228 \\
$C^{\mathrm{rms}}_{l}$ & 1.6735 (0.6038\%) & 1.6826 (0.0604\%) & 1.6827 \\
$f^{*}_{A_y}$      & 0.12 (9.09\%)    & 0.11             & 0.11 \\ \hline
\end{tabular} 

\caption{\label{tab:Tab1mesh} Mesh convergence study results for the elastically mounted rigid inverted foil at $Re = 1000$, $m^{*} = 1$, and $K^*_{\theta} =2.4$}
\end{table*}

In the inverted configuration, a rigid foil with its trailing edge fixed and leading edge free to oscillate is inherently unstable. Consider an undamped, elastically mounted rigid foil with $m^* = 1$, subject to uniform fluid flow at fixed $Re$. As the non-dimensional spring stiffness decreases, the foil exhibits instability dynamics similar to those of a flexible inverted foil \citep{Gurugubelli2019LargePeriodicity, leontini2022dynamics}. In equilibrium, the foil experiences fluid loading as the flow moves around its surface, balanced by the elastic forces of the torsional spring. As stiffness decreases further, it first undergoes static divergence \citep{Kim2013FlappingFlag,Gurugubelli2015Self-inducedFlow,leontini2022dynamics}, and eventually, dynamic instability arises when the foil can no longer dissipate the incoming fluid kinetic energy. These instability regimes provide the basis for generating diverse training data across stiffness values, ensuring that the surrogate model can capture both stable and unstable flapping responses. Capturing these instability transitions in the dataset ensures that the surrogate model learns both the onset and evolution of flapping instabilities.

Distinct flapping regimes emerge as stiffness decreases. Initially, the foil oscillates about a deflected position biased to one side. Further reduction leads to the small-amplitude flapping (SAF) regime, characterized by low-frequency oscillations with a heteroclinic-type orbit \cite{leontini2022dynamics}. As stiffness continues to decrease, the foil transitions into the large-amplitude flapping (LAF) regime, exhibiting symmetric, periodic oscillations with a constant frequency, similar to those of an elastic foil. Beyond this regime, the system ultimately settles into a deflected state. Along with these flapping modes, an asymmetric chaotic flapping regime is also observed between the small- and large-amplitude flapping states and again as the foil transitions from large-amplitude flapping to the deflected position \cite{leontini2022dynamics}. Together, these regimes provide a comprehensive benchmark for training and validating the surrogate model across a wide range of dynamical behaviors.

We next vary the mass ratio $m^*$ while keeping $Re$ and $K^*_{\theta}$ fixed to examine its effect on flapping dynamics. In this work, we consider a constant Reynolds number and spring stiffness of $Re = 1000$, $K^{*}_{\theta} = 2.4 $,  and a damping coefficient $\mathrm{C}_\theta = 0$, with $m^*$ varied in the interval $[0.01, 1]$. For these fluid-structure parameters, the foil undergoes large-amplitude limit cycle oscillations about the stationary position $(y = 0)$, as shown in Fig. \ref{fig:osci}. Beyond $m^* = 1$, for $K^*_{\theta} = 2.4$ and $Re = 1000$, the foil transitions out of the limit cycle LAF regime and into the deflected flapping mode. The choice of $K^{*}_{\theta}$ and $Re$ ensures that the system remains in the LAF regime across all $m^*$ values, as this regime is of interest for energy harvesting \cite{Kim2013FlappingFlag, Gurugubelli2015Self-inducedFlow}. 
Although the present data set is restricted to $Re = 1000$, $K^{*}_{\theta} = 2.4$, this choice reflects the canonical vortex-induced vibration regime of the inverted foils. Future work will extend $\phi$-GNN validation to a parametric sweep across $Re$ and stiffness to test robustness.

\begin{figure}[ht]
\centering\includegraphics[width=0.65\linewidth]{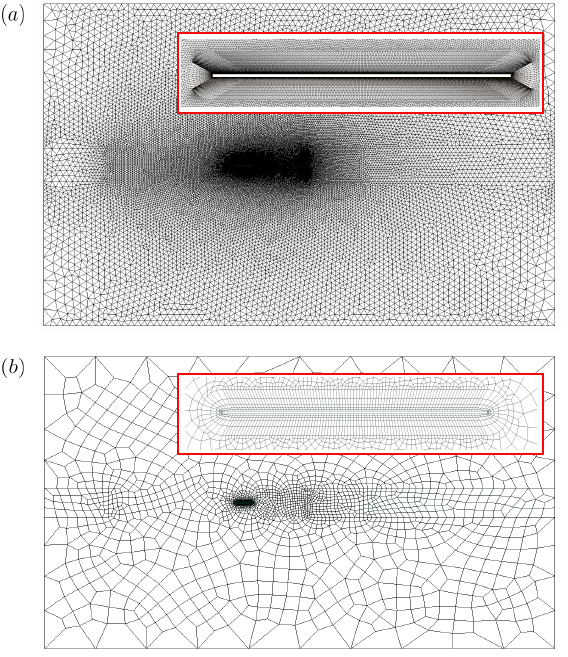}
\caption{(a) Computational mesh used for data generation with a close-up of the boundary layer mesh surrounding the rigid foil, and (b) coarse mesh used for the GNN model}\label{fig:FOMtoGNN}
\end{figure}
\begin{table*}
\centering
\begin{tabular}{lll}
          & FOM mesh &GNN mesh  \\ \hline
       Nodes   & 46075 & 3639 \\      
       $A^{rms}_{y}/L$& 0.5258 & 0.5259 (0.5259$\%$)\\
       $A^{max}_{y}/L$& 0.6796 & 0.6847 (0.7357$\%$)\\
       $\overline{C_{d}}$  & 1.1663 & 1.1757(0.8125$\%$)\\
       $C^{rms}_{l}$  & 1.7181 & 1.72976 (0.6182$\%$)\\ \hline

\end{tabular} 

\caption{\label{tab:Tab2mesh} Simulation results for the elastically mounted rigid inverted foil at $Re = 1000$, $m^{*} = 0.5$, and $K^*_{\theta} =2.4$ on FOM and GNN mesh }
\end{table*}
Within $m^*  \in [0.01, 1]$, the mass ratio is varied from $0.01$–$0.09$ and $0.017$–$0.097$ with uniform increments of $0.01$, and $0.1$-$1$ and $0.17 $-$ 0.97$ with increments of $0.1$. Two additional points at $m^* = 0.83$ and $m^* = 0.86$ are included to capture the sharp increase in the flapping amplitude observed when $m^* > 0.8$ as shown in Fig. \ref{fig:Ap2pc}. This results in a total of 40 cases for the training data set. To evaluate the GNN model, test data are generated by varying $m^*$ from $0.015$-$0.095$ with increments of $0.01$ and $0.15$-$ 0.95$ with increments of $0.1$, yielding 18 cases. A timestep of $\Delta t= 0.02$ is considered for the simulations. This structured dataset enables the GNN to be trained and validated across a wide range of $m^*$ values, capturing both gradual and abrupt changes in flapping response.

\begin{figure}[ht]
\centering
\includegraphics[width=0.8\linewidth]{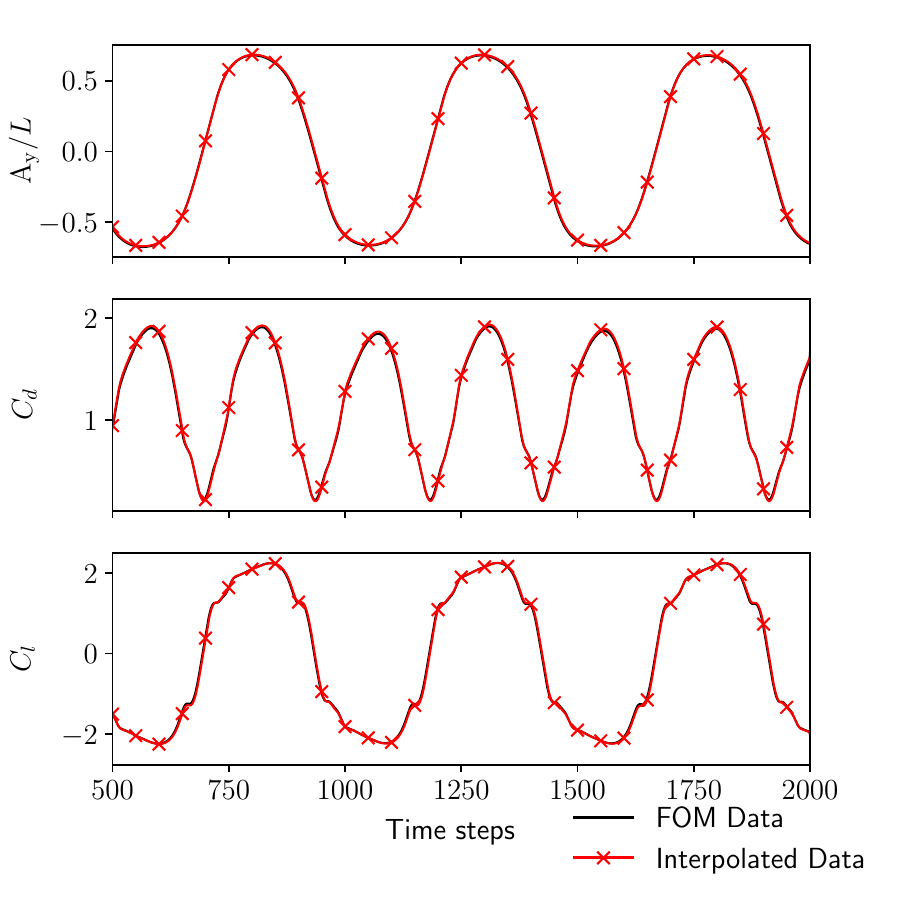}
\caption{Transverse tip displacement, lift, and drag coefficients calculated from the fine finite element mesh and interpolated coarse GNN grid 
 at $m^* = 0.5$, $Re = 1000$ and $K^*_{\theta} = 2.4$}\label{fig:Interp}
\end{figure}

The training and evaluation of the GNN using features derived from a refined finite element mesh are computationally expensive. To reduce complexity, the high-fidelity simulation data are interpolated onto a coarser mesh suitable for GNN training, as shown in Fig. \ref{fig:FOMtoGNN}. The interpolation is performed using the Clough–Tocher 2D interpolation scheme, available in the SciPy package\citep{virtanen2020scipy}. Since mesh resolution directly impacts GNN prediction accuracy, selecting an optimal mesh is crucial. In this study, a coarse GNN mesh with $3,639$ nodes is selected, as shown in Fig. \ref{fig:FOMtoGNN}b. The coarse grid is generated with a refined mesh resolution in the boundary layer and near wake region, and a sparser resolution surrounding the less sensitive areas away from the foil. Figure \ref{fig:Interp} presents the comparison of the time series evolution of the transverse tip displacement, lift, and drag coefficients calculated using the full order mesh and the coarse GNN mesh for the rigid inverted foil at $Re=1000$, $m^* = 0.5$, and $K^*_{\theta}= 2.4$. The rigid foil's response characteristics, particularly the transverse tip displacement and coefficients of lift and drag forces, are used as metrics to evaluate the interpolation of the high-fidelity data onto the GNN grid as listed in Table \ref{tab:Tab2mesh}.  As observed in Fig. \ref{fig:Interp} and Table \ref{tab:Tab2mesh}, the selected coarse mesh reproduces the flapping response characteristics with interpolation errors below 1$\%$. This confirms that the coarse mesh provides an efficient yet accurate representation for training the GNN surrogate.

For each $m^*$ in the dataset, $2,500$ continuous time steps are recorded from the ground-truth simulations once the system reaches a statistically stationary state and are interpolated onto the coarse grid to construct the ground-truth dataset. For the $m^*$ values in the training dataset, 2,048 continuous samples of spatial data, starting from the time step $n_h + 1$ are used to train the GNN model, while the remaining samples are used for cross-validation. Here, $n_h$ denotes the number of history steps used to train the solid/mesh prediction sub-network. In this work, information from $n_h=9$ history time steps is provided to the mesh sub-network. An internal sensitivity analysis was conducted on the $n_h$, varying this parameter from $7$ to $11$. Based on this study, we determined that $n_h=9$ time steps are sufficient to produce stable predictions of the mesh states. Both increasing and decreasing $n_h$ lead to reduced accuracy and compromise the network’s stability for long-duration predictions. The number of low-order retained modes is fixed at $n_r =2$ across all $m^*$ cases within the training and testing datasets. To evaluate the network’s forecasting performance, 1,501 time steps are collected for each $m^*$ in the test dataset, also beginning from $n_h + 1$. This dataset structure ensures balanced training and testing of the GNN across different mass ratios while capturing both short- and long-horizon dynamics.

\subsection{Network Architecture and Training}
In this work, we employ an architecture similar to that of Gao \textit{et al.}\cite{gao2024predicting}, implemented in the PyTorch library \cite{paszke2019pytorch}. For mesh displacement prediction, the MLPs consist of three hidden layers, each with 512 neurons. For the hypergraph sub-network, the encoders, decoders, node, and element update functions within each message passing layer each utilize MLPs with two hidden layers, each with a width of 128 neurons. The GNN architecture comprises 15 message passing layers, ensuring adequate receptive field coverage and enabling the model to capture multi-element interactions observed in complex fluid-structure interaction problems \citep{gao2024finite}. It is important to note that the hyperparameters of hypergraph width and depth are fixed and not tuned following the choices of \citep{pfaff2020learning} as well as our work \cite{gao2024finite,gao2024predicting, gao2024towards}.  A sinusoidal activation function \citep{sitzmann2020implicit} is applied across all the layers for both the sub-networks. The mean aggregation function is used for both node and element updates, and a gather–scatter scheme is employed to implement the message-passing layers. Building on our earlier work \citep{gao2024finite,gao2024predicting}, no normalization layers are used to avoid training instabilities.

The mesh/solid and fluid sub-networks are trained separately on a single Nvidia A100 GPU. The input and output features of the training dataset are normalized using a min-max normalization with a range of $[-1, 1]$, except for $\mathrm{cos \theta}_{s,i}$ in $e_{s,i}$, which are inherently within this range. The cross-validation and test datasets are normalized using the statistics computed from the training dataset. An Adam optimizer \cite{kingma2014adam} with $\beta_1 = 0.9$ and $\beta_2 = 0.999$ is used to train both the sub-networks. 
The mesh prediction sub-network is trained with a batch size of $256$ over $210$ epochs. A hybrid cosine-to-exponential scheduler is applied for training the mesh prediction sub-network, where the learning rate gradually decreases from $10^{-4}$ to $1\times10^{-6}$. Gradient clipping \citep{pascanu2013difficulty}with a threshold of $0.015$ is applied to the network's parameters to prevent exploding gradients. The clipping threshold is applied such that if the total gradient norm exceeds the threshold value, the gradients are scaled proportionally to maintain stability. Early stopping with a patience of $30$ epochs was also applied to prevent overfitting, wherein the training is stopped if the validation loss does not improve by $4.5\times 10^{-6}$ during this period.
The fluid sub-network is trained with a batch size of 4 with a learning rate schedule spanning 50 epochs, including 5 warm-up epochs. During the warm-up period, the learning rate increases per iteration from $10^{-6}$ to $10^{-4}$. After warm-up, the learning rate is gradually decreased from $10^{-4}$ to $10^{-6}$ for the remaining 45 epochs using the same hybrid cosine-to-exponential scheduler as the mesh prediction sub-network.

An adaptive smooth $L_1$ loss function is employed during training \cite{gao2024predicting}, defined as:
\begin{align}
    \mathcal{L}^{L_1}_i\left(\psi_i, \hat{\psi}_i\right)= \begin{cases}\left(\psi_i-\hat{\psi}_i\right)^2 / 2 \beta, & \text { if }\left|\psi_i-\hat{\psi}_i\right|<\beta \\ \left|\psi_i-\hat{\psi}_i\right|-\beta / 2, & \text { if }\left|\psi_i-\hat{\psi}_i\right| \geq \beta\end{cases}
\end{align}
with the non-negative control parameter $\beta$ adapted in every training iteration \cite{pfaff2020learning,gao2024predicting}
\begin{align}
    \beta^2 \leftarrow\left(1-\frac{1}{N_b}\right) \beta^2+\frac{1}{N_b} \min \left\{\beta^2, \operatorname{MSE}\left(\psi_{\text {batch }}, \hat{\psi}_{\text {batch }}\right)\right\},
\end{align}
where $\psi$ and $\hat{\psi}$ denote the ground truth output and network predicted output for a given batch, the subscript $(\cdot)_i$ denotes the entry-by-entry calculation, $N_b$ represents the number of training epochs, and MSE denotes the mean squared error. Noise is added during the training of both sub-networks following the scheme devised by Gao \textit{et al.}\cite{gao2024finite}. The standard deviation of the noise is set to $0.1$ for both the mesh and flow prediction networks, with an over-correction factor $\omega =1.2$.  A summary of the architectures for each sub-network and the training schemes is provided in Table \ref{tab:Tab1Sum}. 
\begin{table*}

\begin{tabular}{lll}
         Component & Mesh/Solid Sub-network & Flow Sub-network \\ \hline
         Batch size & $256$ & $4$ \\   
         Warmup epochs&$0$ &$5$\\
         Training epochs&$210$ &$45$\\
       Optimizer& \multicolumn{2}{c}{Adam}\\
       Learning rate (min, max) &  \multicolumn{2}{c} {$10^{-6}, 10^{-4}$}\\
        Gradient clipping &Global norm threshold of $0.1$&-\\
        Gaussian noise &  \multicolumn{2}{c} {$ \sigma=0.1, \omega=1.2$} \\
        Network Architecture& $3$-layer MLP & $2$-layer MLP \\
        &$512$ neuron width&$128$ neuron width\\
        && $15$ message passing layers\\\hline

\end{tabular} 

\caption{\label{tab:Tab1Sum}Summary of architecture and training settings for the mesh and flow state prediction sub-networks}
\end{table*}
The mesh prediction network trains in approximately $1$ second per epoch, while the flow prediction network requires $2060$ seconds per epoch. Once trained, the hypergraph network auto-regressively generates rollout predictions of the system state over $1500$ time steps starting from time step $h+1$ for each $m^*$ value in the test set. These predictions are obtained using only a fraction of the time taken by the high-fidelity finite element simulations.  The training and inference times of the framework are summarized in Table \ref{tab:Tab1Time}. Overall, the framework achieves more than two orders of magnitude speed up compared to the ground truth simulations performed on the fine finite element mesh. In the following subsections, we evaluate the hypergraph network's ability in capturing the complex, large amplitude flapping dynamics of inverted foils by comparing its predictions with the corresponding ground truth values. 
\begin{table*}
\centering
\begin{tabular}{lll}
Metric & Mesh sub-network & Flow sub-network \\ \hline
Train time (s/epoch)    & $<1$           & $\approx 2060$ \\ 
Inference speed (ms/step) & \multicolumn{2}{c}{$\approx 10.0$} \\ \hline 
\end{tabular}
\caption{\label{tab:Tab1Time}Training and inference (including post-processing time) time of the framework}
\end{table*}

\section{Results and Discussion} \label{Sect:4}
In this section, we present the performance of the proposed $\phi$-GNN surrogate model in predicting the coupled flapping dynamics of inverted foils. The results are compared against high-fidelity simulations to assess both qualitative and quantitative accuracy. We first examine the ability of the network to reproduce the foil’s oscillatory response across a range of mass ratios, followed by an evaluation of near-wake flow structures and fluid loading characteristics. Finally, we discuss the implications of these findings for energy harvesting and outline the model’s strengths and limitations.

\subsection{Flapping Response}\label{subsect: Response}
To evaluate the network's ability to predict the foil's fluid-structure interaction dynamics, we examine how well the model captures the variation in flapping amplitude of the foil across different mass ratios. We consider two metrics for analysing the structure's response: the normalized peak transverse tip displacement (leading edge) $A^{peak}_{y}/2L$ and the r.m.s. of transverse tip displacement defined as $A^{rms}_{\mathrm{y}}/L=\left(\sqrt{\frac{1}{n}\Sigma_{t}(A_{\mathrm{y}}-\overline{A_{\mathrm{y}}}})^{2}\right)/L$ as metrics to evaluate the GNN framework. Where $A_{\mathrm{y}}$  and $\overline{A_{\mathrm{y}}}$ represent the transverse tip displacement and the mean transverse displacement,  respectively. To quantitatively assess the model's forecasting performance, we compute the coefficient of determination $(R^2)$, defined as the square of the Pearson correlation coefficient between the predicted and ground truth values, along with the absolute percentage relative error. 
\begin{figure}[ht]
\centering\includegraphics[width=0.65\linewidth]{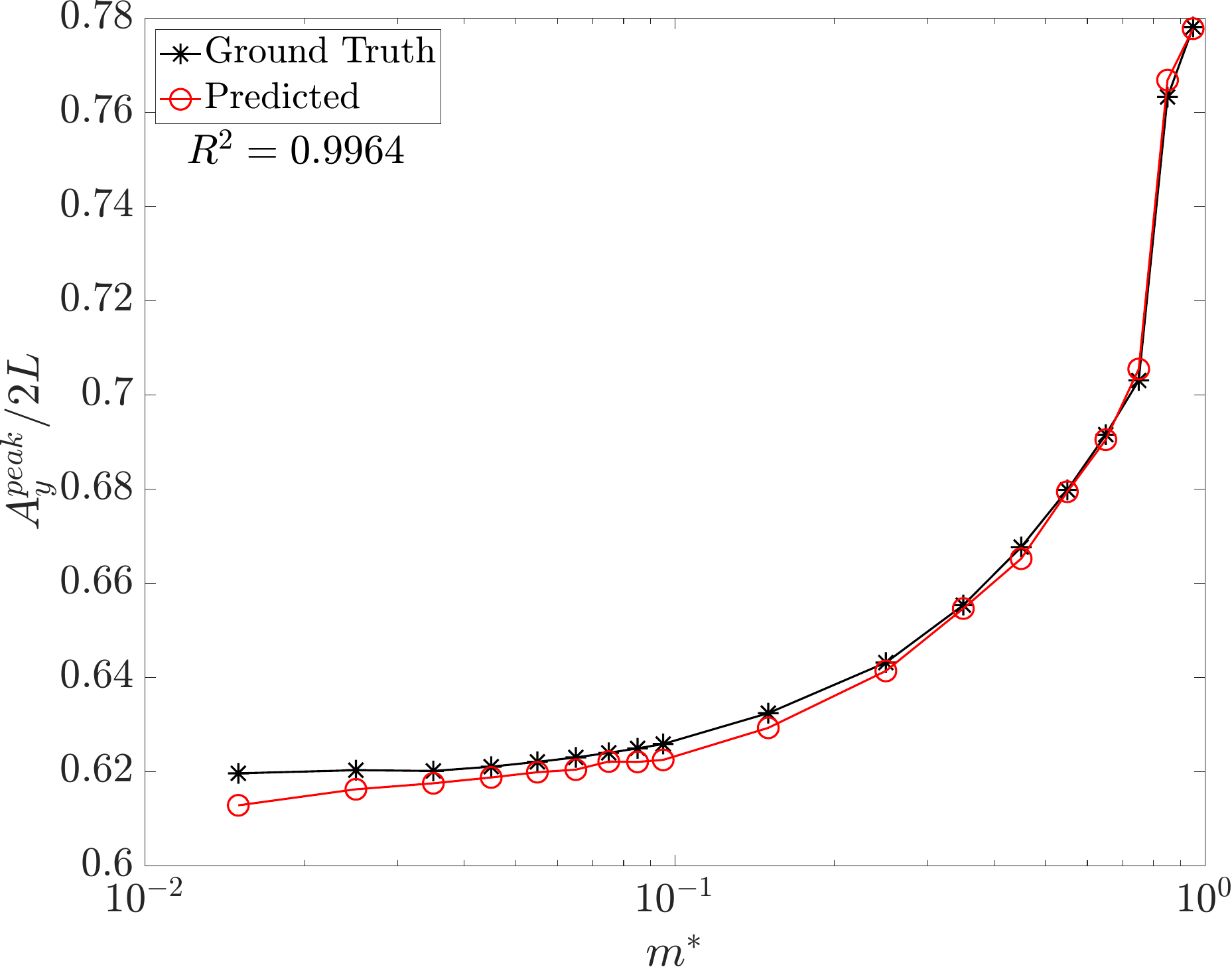}
\caption{Variation of average peak to peak transverse tip displacement as a function of $m^*$}\label{fig:AP2P}
\end{figure}

\begin{figure}[ht]
\centering\includegraphics[width=0.55\linewidth]{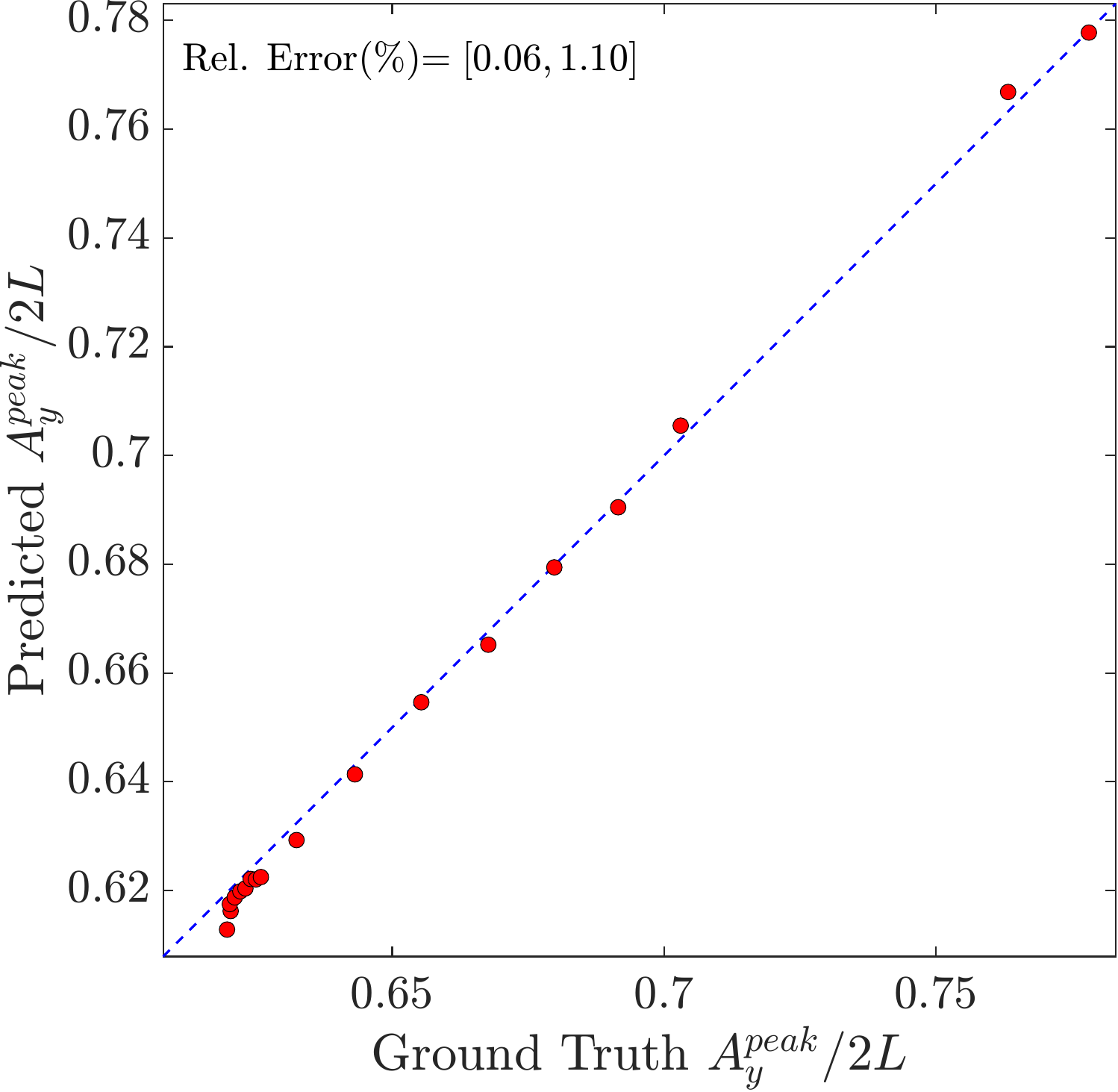}
\caption{Comparison between predicted and ground truth values for the average peak to peak transverse tip displacement of the foil}\label{fig:Ap2pc}
\end{figure}

\begin{figure}[ht]
\centering\includegraphics[width=0.65\linewidth]{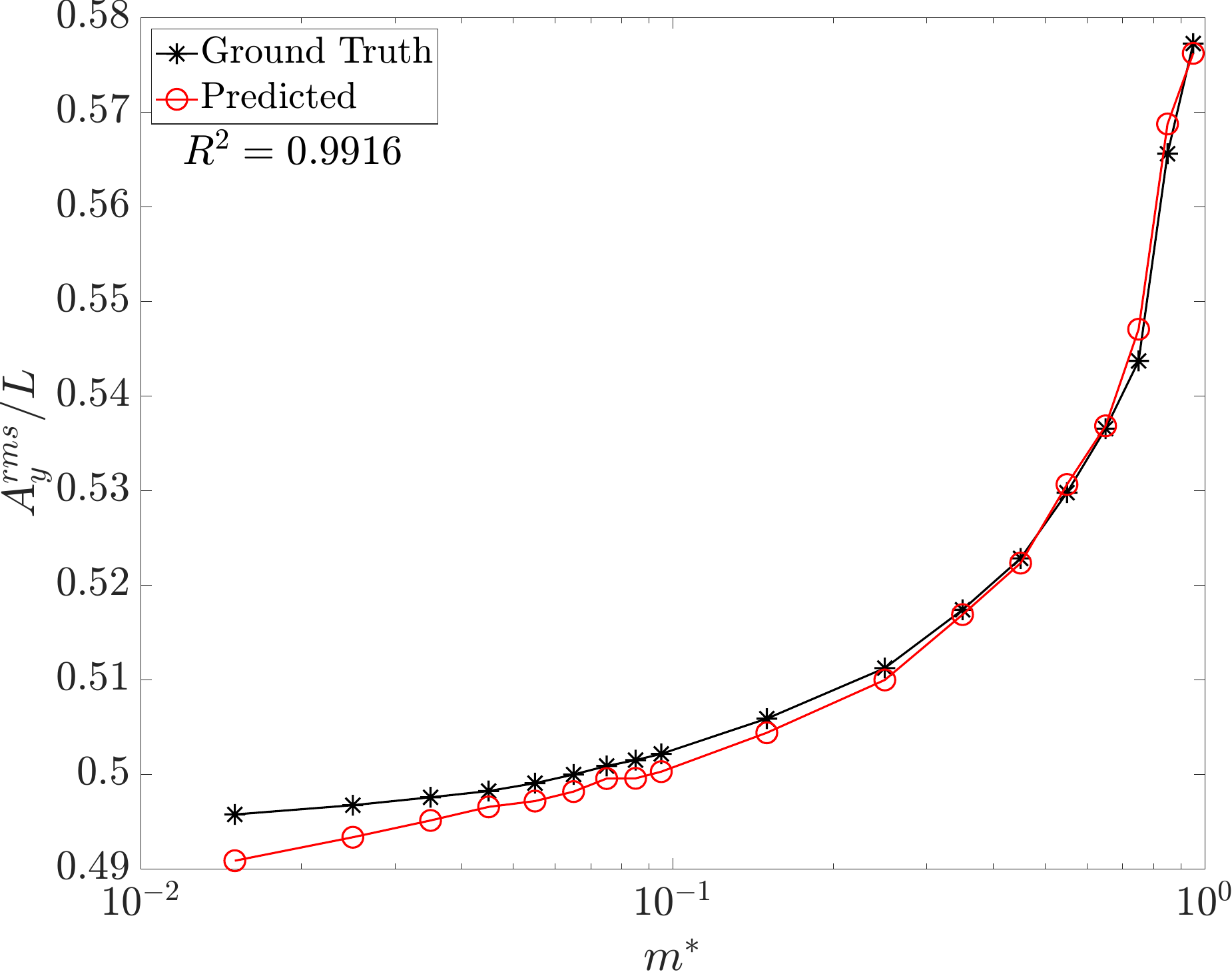}
\caption{Variation of r.m.s of transverse tip displacement as a function of $m^*$}\label{fig:Arms}
\end{figure}

\begin{figure}[ht]
\centering\includegraphics[width=0.55\linewidth]{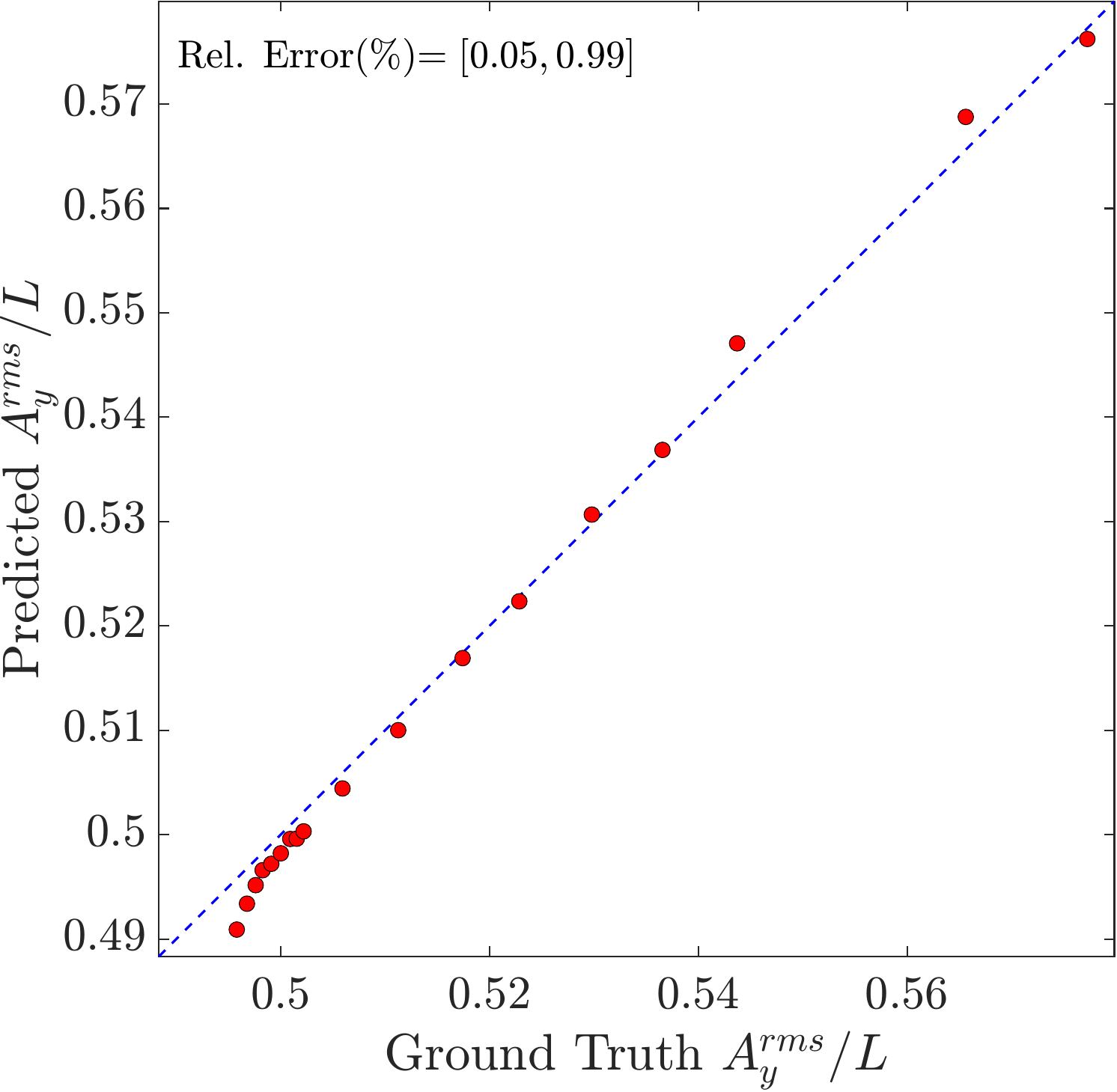}
\caption{Comparison between predicted and ground truth values for the normalized r.m.s of transverse tip displacement}\label{fig:Armsc}
\end{figure}
Figures \ref{fig:AP2P} and \ref{fig:Arms} present the variation of $A^{peak}_{y}/2L$ and $A^{rms}_{\mathrm{y}}/L$ as a function of $m^*$ for both GNN predictions and ground truth high fidelity data. In the LAF regime, for a given stiffness and $Re$, the foil's oscillation amplitude increases with increasing $m^{*}$ as captured by the high-fidelity simulations. At high $m^*$ values of $O(1)$, the foil is relatively heavy compared to the surrounding fluid. Consequently, structural inertia dominates the response, leading to higher flapping amplitudes. As $m^*$ decreases to $O(10^{-2})$, the foil becomes lighter with respect to the fluid, reducing the influence of structural inertia. As a result, the foil undergoes oscillations with lower amplitudes when interacting with the surrounding fluid. This trend is consistent with earlier experimental \citep{Kim2013FlappingFlag, tavallaeinejad2021dynamics} and numerical studies \citep{Gurugubelli2015Self-inducedFlow, Goza2018GlobalFlapping, parekh2025wake} on flexible inverted foils, indicating that the underlying physics is common across both rigid and flexible systems. This trend establishes the physical baseline against which the GNN predictions are evaluated.

The hypergraph framework successfully reproduces this behavior, closely matching the ground truth values across the entire $m^*$ range. It can capture the increase in the flapping amplitude with $m^*$, from the gradual increase up to $m^* \leq 0.7$ and the jump in amplitude for $m^*>0.7$, as shown in Figures \ref{fig:AP2P} and \ref{fig:Arms}. The model's predictions show excellent quantitative agreement with the ground truth values, with $R^2 = 0.9916$ and a maximum relative error of $1.1\%$ for the peak tip displacement and $R^2 = 0.9964$ and a maximum relative error of $0.99\%$ for the RMS tip displacement. These results highlight the $\phi$-GNN’s accuracy in capturing amplitude variations across a wide mass-ratio spectrum.

Figures \ref{fig:Ap2pc} and \ref{fig:Armsc} illustrate the point-wise comparison between the predicted and ground truth values of $A^{peak}_{y}/2L$ and $A^{rms}_{\mathrm{y}}/L$  along with the corresponding absolute relative errors bounds. The close clustering of the points along the parity line illustrates the model's consistency and the narrow error bounds confirm that the deviations remain small across the entire $m^*$ range, together with Figures \ref{fig:AP2P} and \ref{fig:Arms}, these results confirm that the network captures the underlying physical behavior while maintaining a high degree of numerical accuracy.

\begin{figure}[ht]
\centering\includegraphics[width=0.65\linewidth]{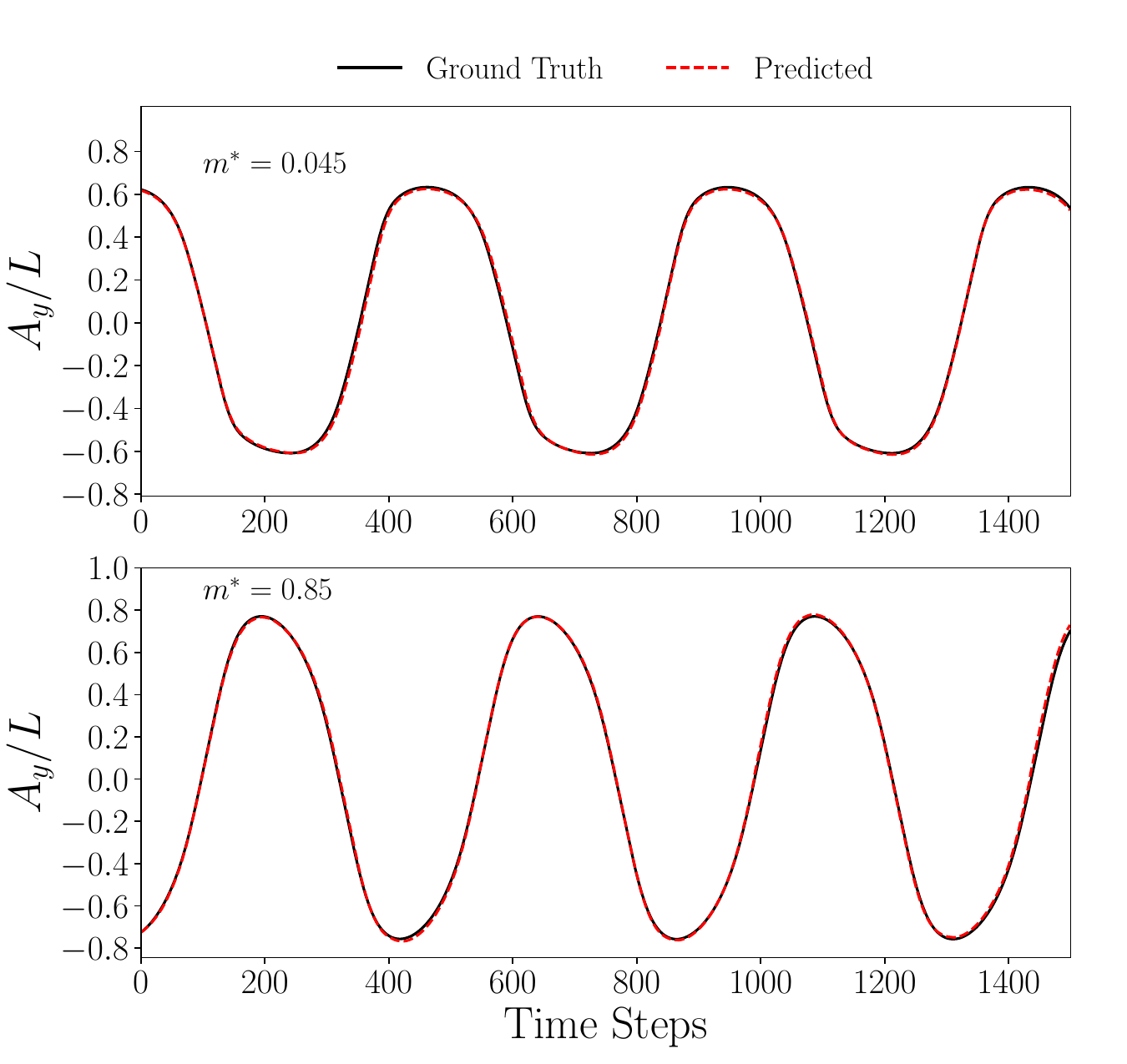}
\caption{Transverse tip displacement calculated from the predicted vs the ground truth
system states for test dataset at $m^* = 0.045$ (top), $m^* = 0.85$ (bottom)  }\label{fig:traj}
\end{figure}

Figure \ref{fig:traj} presents the time histories of the foil's transverse tip displacement $A_y/L$ at test mass ratios $m^*=0.045$ and $m^*=0.85$. These results show that the model generates stable roll-out predictions of the flapping response at both extremes of the amplitude with high accuracy, as demonstrated by the close agreement between the ground truth and predicted trajectories.  Similar to the response shown in Fig. \ref{fig:traj}, the foil undergoes symmetric oscillations about the zero-flow position for all cases in the considered mass ratio range. Although only two cases are shown here, similar levels of accuracy in rollout predictions are achieved across all test cases. Overall, these results demonstrate that, in addition to capturing the underlying physical behavior of the structural response, the network generates stable predictions of the mesh state and accurately resolves the temporal evolution of the response.

\subsection{Flow Field Characteristics}
\begin{figure*}[ht]
    \centering
    \includegraphics[width=0.95\textwidth]{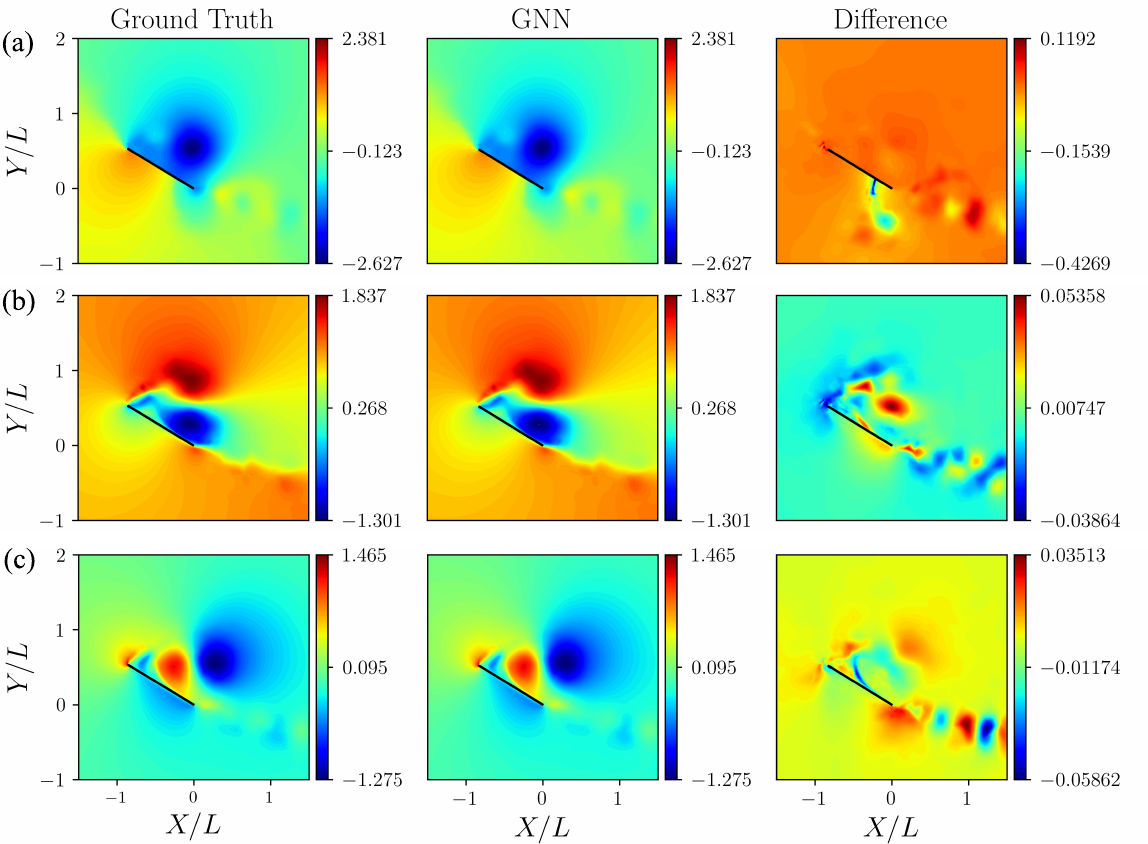} 
    \caption{Comparison between predicted and ground truth contours of (a) normalized pressure $(2p/\rho_{\mathrm{f}}U^2_{\infty} )$, (b) streamwise
velocity $\mathrm{u_x}$, and (c) cross-flow velocity $\mathrm{u_y}$ for the rigid foil at a test $m^*=0.045$; fields are shown at roll-out time
step 1500. }
    \label{fig:flowfield}
\end{figure*}
\begin{figure*}[ht]
    \centering
    \includegraphics[width=0.95\textwidth]{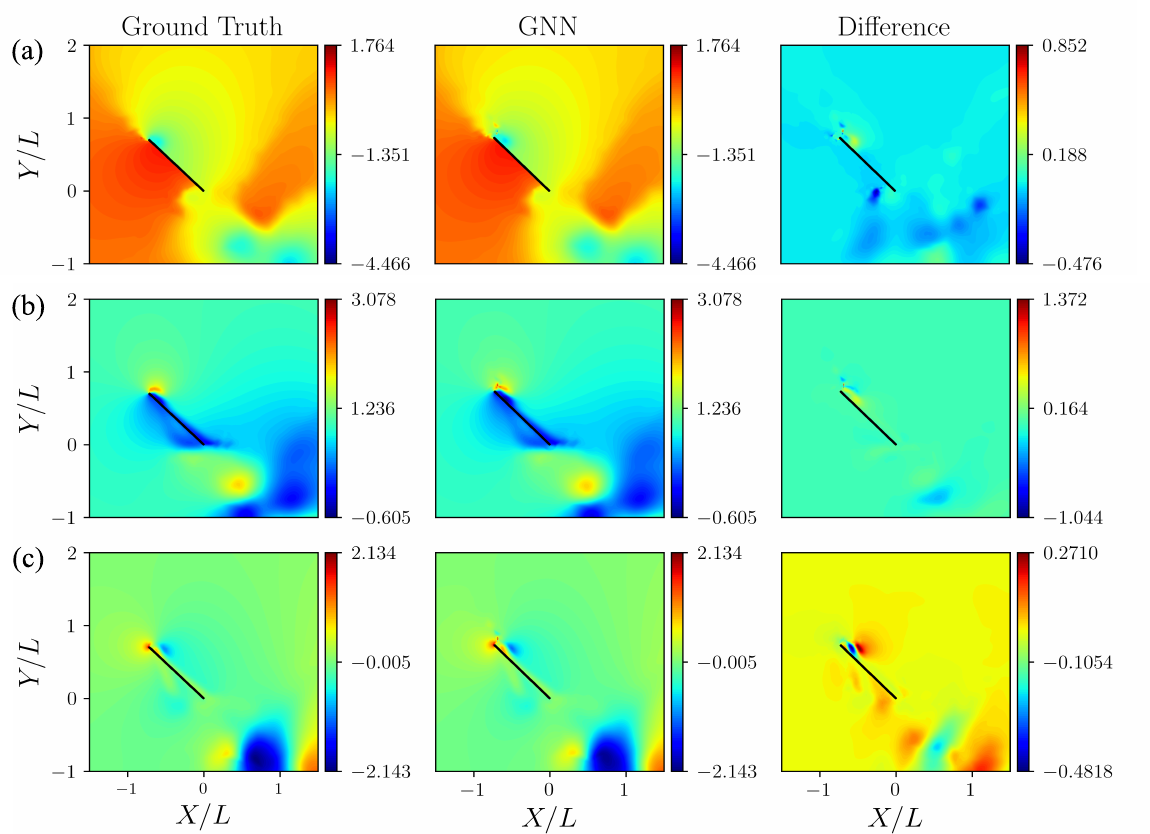} 
    \caption{Comparison between predicted and ground truth contours of (a) normalized pressure $(2p/\rho_{\mathrm{f}}U^2_{\infty} )$, (b) streamwise
velocity $\mathrm{u_x}$, and (c) cross-flow velocity $\mathrm{u_y}$ for the rigid foil at a test $m^*=0.85$; fields are shown at roll-out time
step 1500. }
    \label{fig:flowfield2}
\end{figure*}
So far, we have observed that the hypergraph framework can accurately capture the oscillatory response of the inverted foil when subject to steady uniform flow across the entire range of $m^*$.Next, we evaluate how the model predicts the evolution of flow field characteristics. Figures \ref{fig:flowfield} and \ref{fig:flowfield2} present the comparison between the predicted and ground truth flow fields of the normalized pressure $2p/\rho_{\mathrm{f}}U^2_{\infty}$, and velocity $(\mathrm{u_x,u_y})$ for test mass ratios $m^*=0.045$ and $m^*=0.85$, respectively. The predicted flow fields illustrate that the complex flow patterns and wake features observed in the ground truth are qualitatively reproduced by the $\phi$-GNN. The spatial structure and flow field magnitude show good agreement with the ground truth data for both $m^*$. The prediction errors are concentrated in the near wake of the foil, where interaction between the strong unsteady vortices formed and shed at the foil's leading and trailing edges results in highly non-linear flow dynamics. In this region, the fluctuating strength of the flow field is slightly over-predicted. Despite these localized discrepancies, the predicted fields effectively capture the key dynamic features of the flow.

\begin{figure}[ht]
\centering
\includegraphics[width=0.65\linewidth]{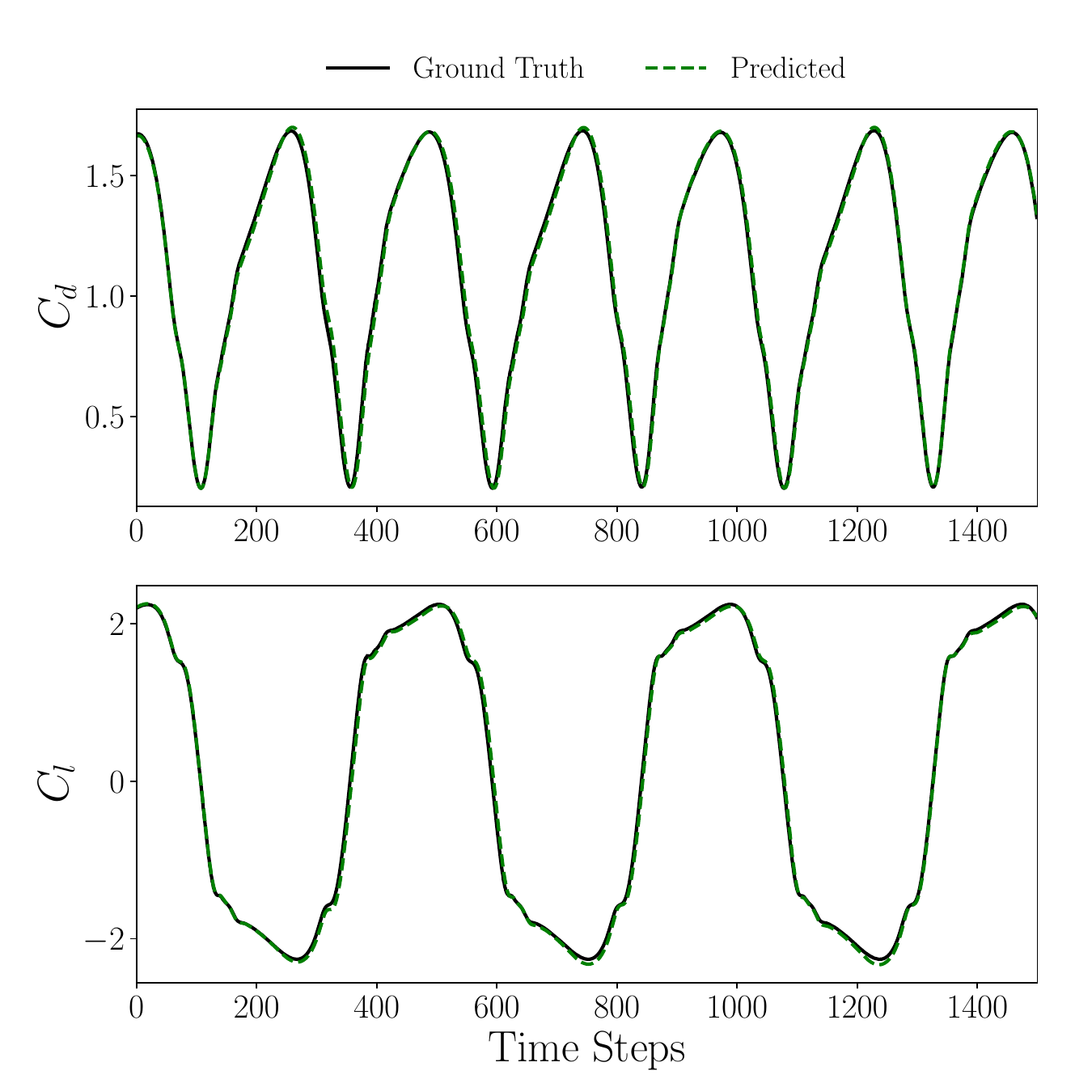}
\caption{Coefficients of drag and lift calculated from predicted vs ground truth system states for the test dataset at $m^*= 0.045$}\label{fig:clcdcomp1}
\end{figure}

\begin{figure}[ht]
\centering
\includegraphics[width=0.65\linewidth]{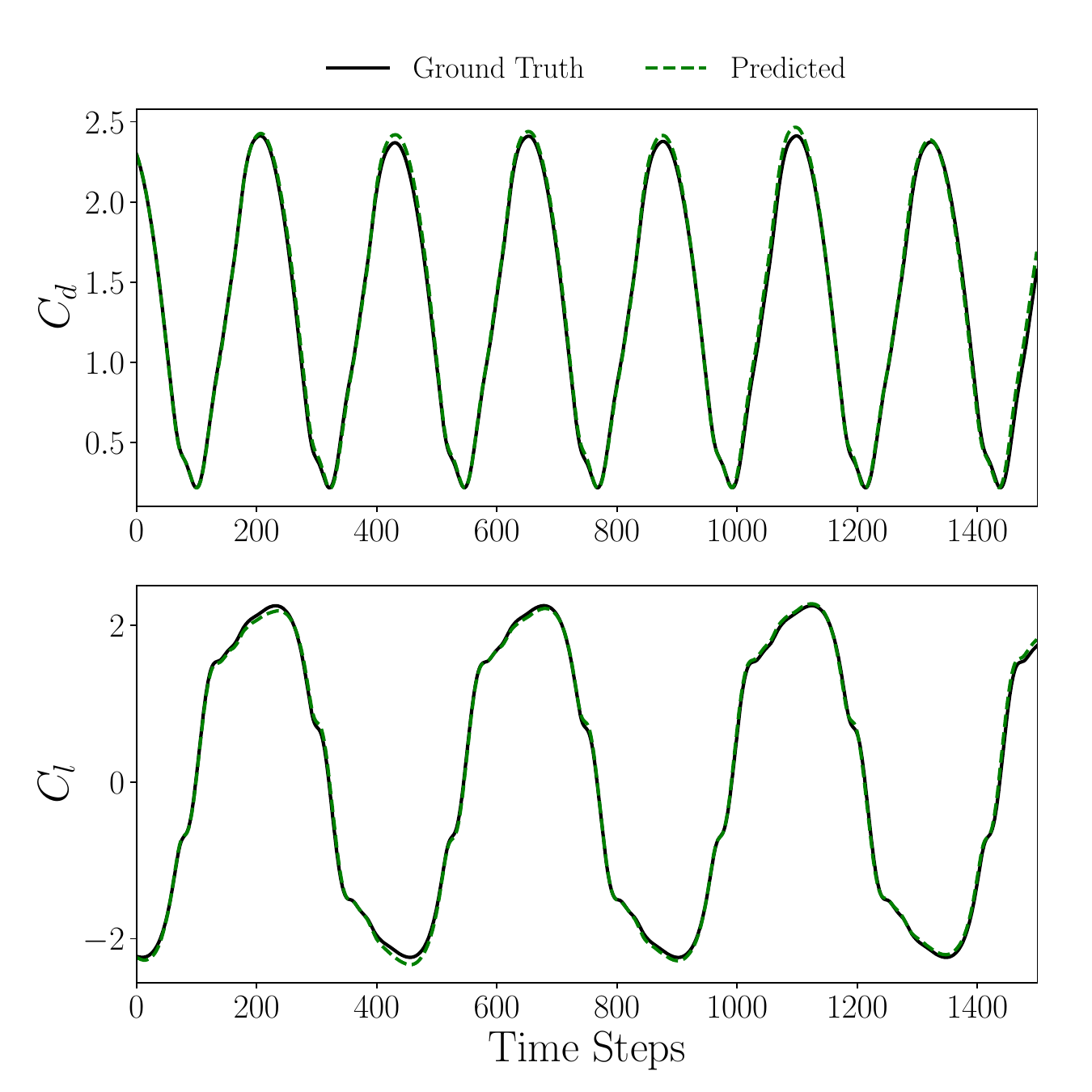}
\caption{Coefficients of drag and lift calculated from predicted vs ground truth system states for the test dataset at $m^*= 0.85$}\label{fig:clcdcomp2}
\end{figure}

Next, we quantitatively evaluate how the model predicts the flow field  by extracting the lift coefficient $C_l$ and drag coefficient $C_d$ on the foil from the system state and comparing them with the ground truth values. The fluid loading on the foil is calculated by integrating the Cauchy stress tensor $\sigma^{\mathrm{f}}$ as defined in Eq. (\ref{eq:Cauchy}) from the first layer of elements away from the solid at the fluid-structure interface.
\begin{align}
    & C_l=\frac{1}{\frac{1}{2} \rho^{\mathrm{f}}\left(\mathrm{U}_{\infty}\right)^2 L} \int_{{\Gamma}^{\mathrm {fs }}}\left(\boldsymbol{\sigma}^{\mathrm{f}} \cdot \mathbf{n}\right) \cdot \mathbf{n}_y d {\Gamma}, \\
& C_d=\frac{1}{\frac{1}{2} \rho^{\mathrm{f}}\left(\mathrm{U}_{\infty}\right)^2 L} \int_{{\Gamma^{\mathrm{f s}}}}\left(\boldsymbol{\sigma}^{\mathrm{f}} \cdot \mathbf{n}\right) \cdot \mathbf{n}_x d {\Gamma},
\end{align}
The lift and drag force coefficients calculated from the ground truth and the predicted system states at two representative cases: a high $ m^* = 0.85$ and low $m^*  = 0.045$ as shown in Figs. \ref{fig:clcdcomp1} and \ref{fig:clcdcomp2}. These results demonstrate that the fluid forces calculated from the predicted states are accurate, and the network can accurately capture the complex characteristics of the flow field around the foil surface. Minor discrepancies are observed in the drag coefficient predictions as observed in Fig. \ref{fig:clcdcomp2}; however, the temporal evolution of the dominant flow features of the fluid loading on the foil is accurately reproduced by the hypergraph framework.  

\begin{figure}[ht]
\centering\includegraphics[width=0.65\linewidth]{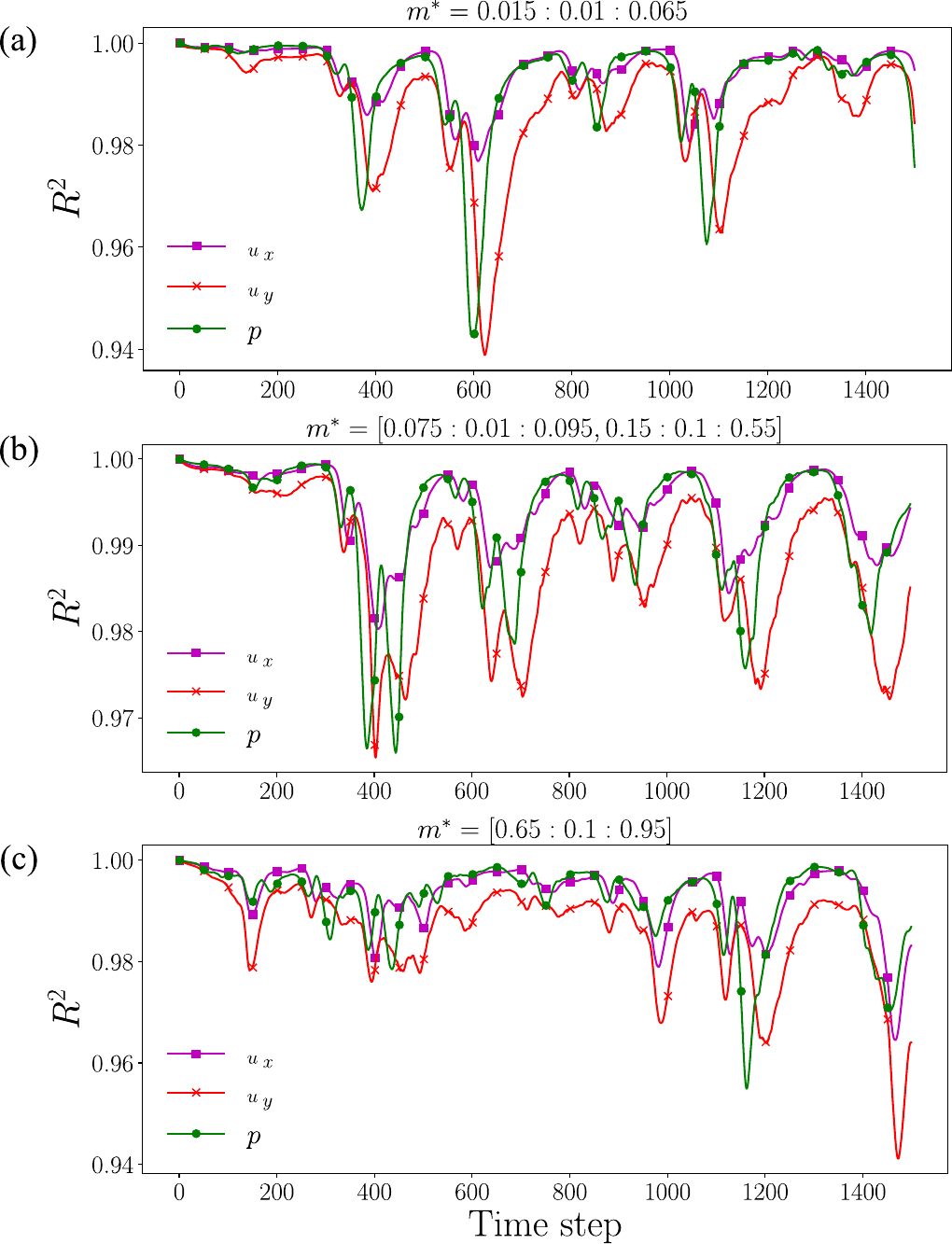}
\caption{Mean coefficient of determination between ground truth and predicted flow fields at every prediction time step at (a) $m^* = [0.015:0.01:0.065]$, (b) $m^* = [0.075:0.01:0.095, 0.15:0.1:0.55]$, (c) $m^* = [0.65:0.1:0.95]$}\label{fig:R2}
\end{figure}

\begin{figure*}[ht]
    \centering
    \includegraphics[width=0.95\textwidth]{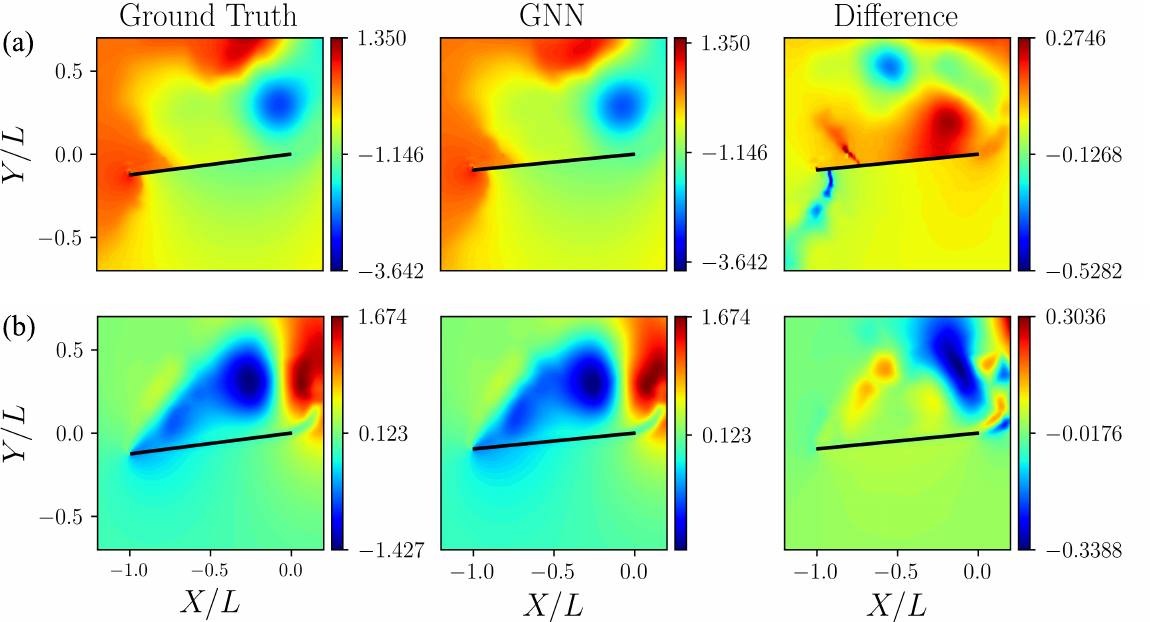} 
    \caption{Comparison between predicted and ground truth contours of (a) normalized pressure $(2p/\rho_{\mathrm{f}}U^2_{\infty} )$ and (b) cross-flow velocity $\mathrm{u_y}$ for the rigid foil at a test $m^*=0.045$; fields are shown at roll-out time step 600. }
    \label{fig:flowfield3}
\end{figure*}

\begin{figure}[ht]
\centering\includegraphics[width=0.65\linewidth]{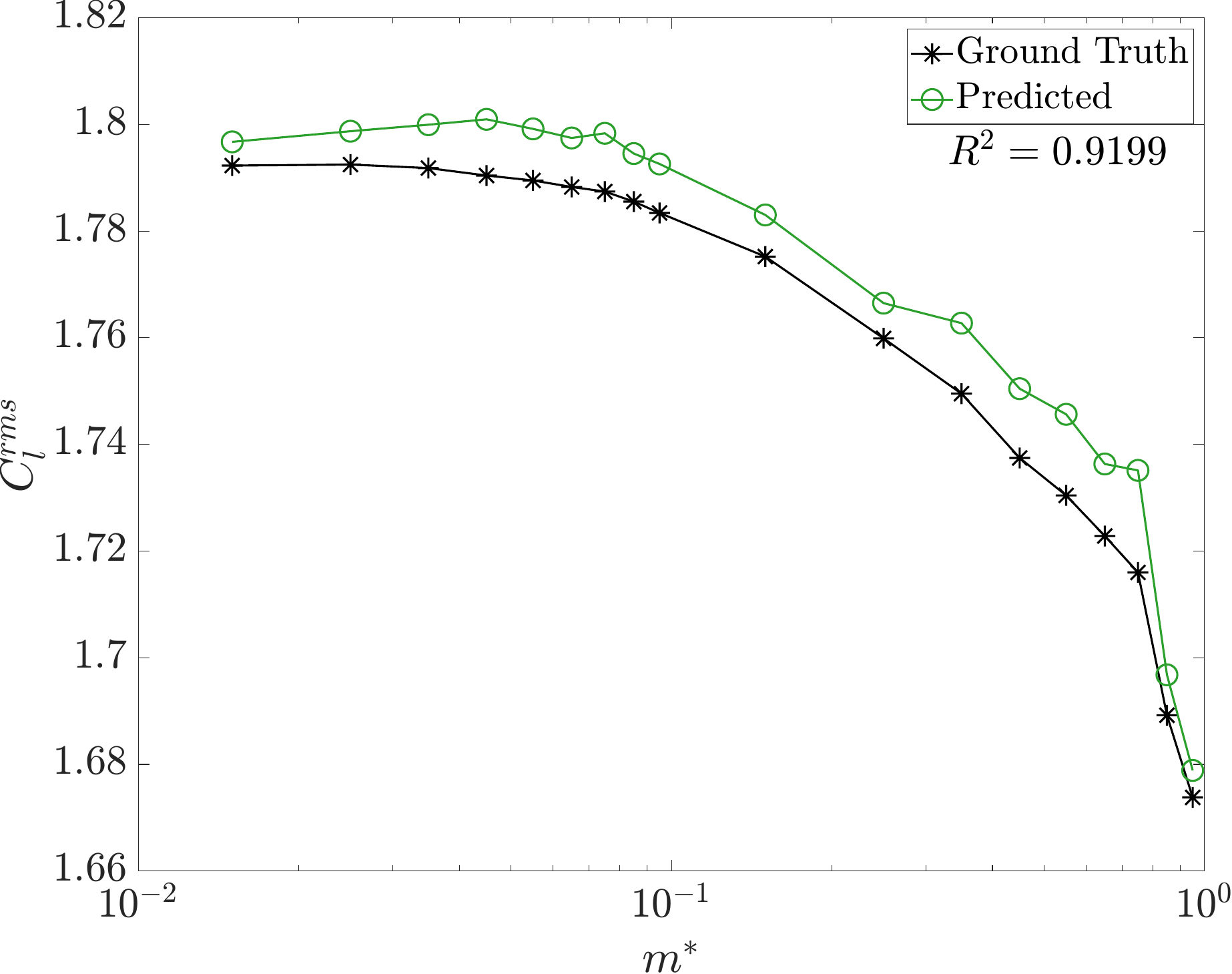}
\caption{Variation of fluctuating root mean square of the coefficient of lift force as a function of $m^*$}\label{fig:Clrms}
\end{figure}

\begin{figure}[ht]
\centering\includegraphics[width=0.55\linewidth]{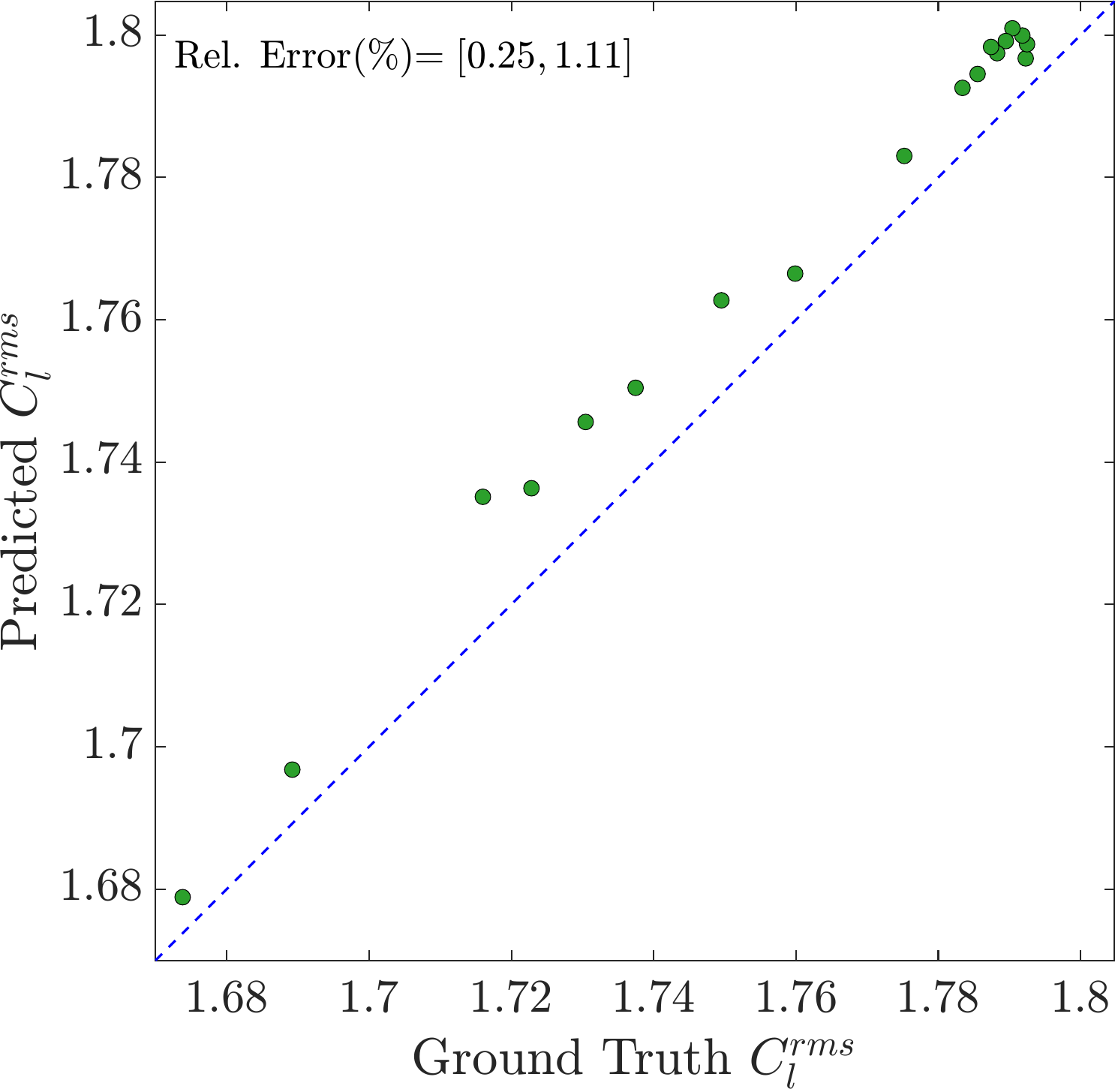}
\caption{Comparison between predicted and ground truth values for the root mean square of the coefficient of lift force on the foil}\label{fig:Clrmsc}
\end{figure}

\begin{figure}[ht]
\centering\includegraphics[width=0.65\linewidth]{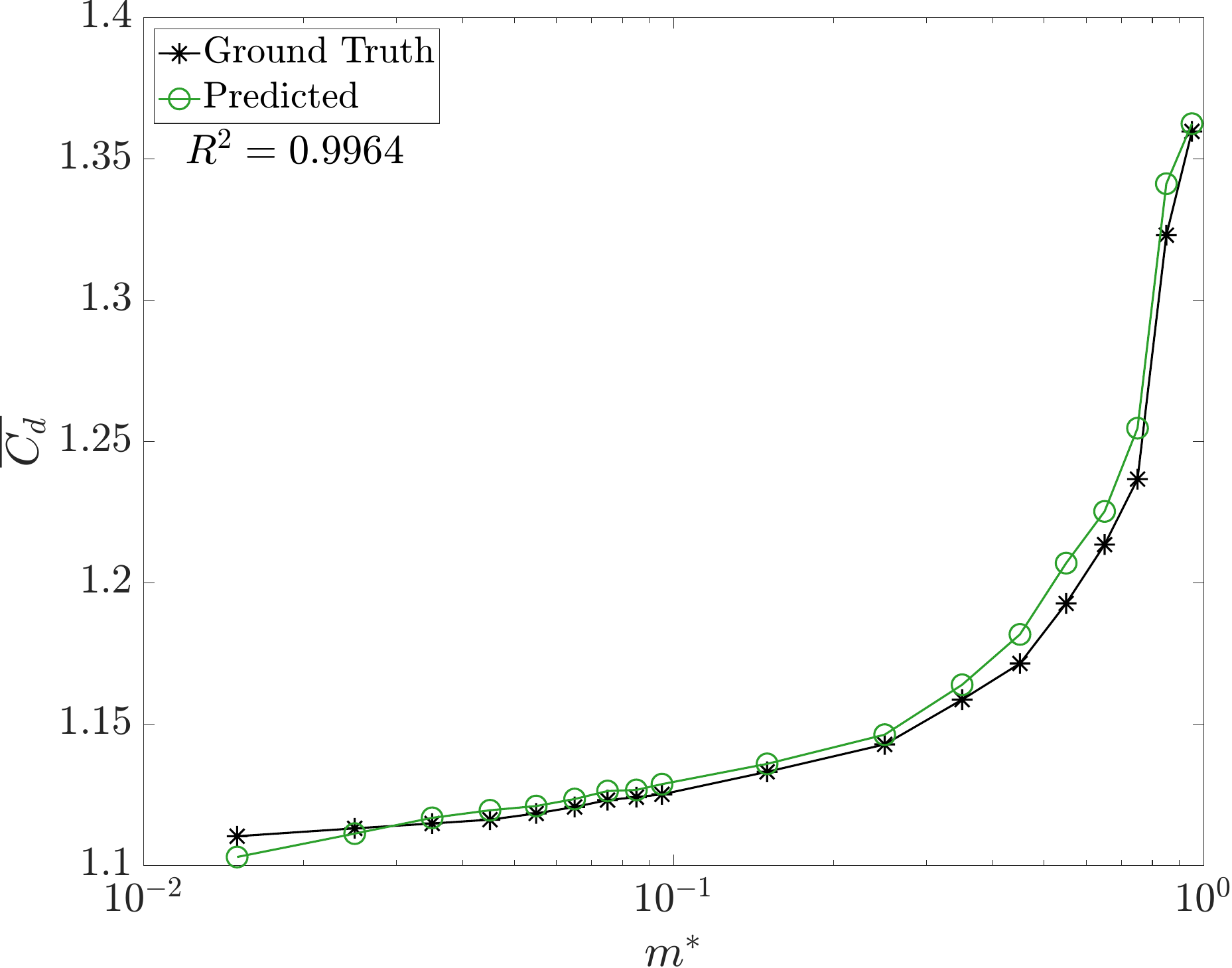}
\caption{Variation of mean coefficient of drag force as a function of $m^*$}\label{fig:cdmean}
\end{figure}

\begin{figure}[ht]
\centering\includegraphics[width=0.55\linewidth]{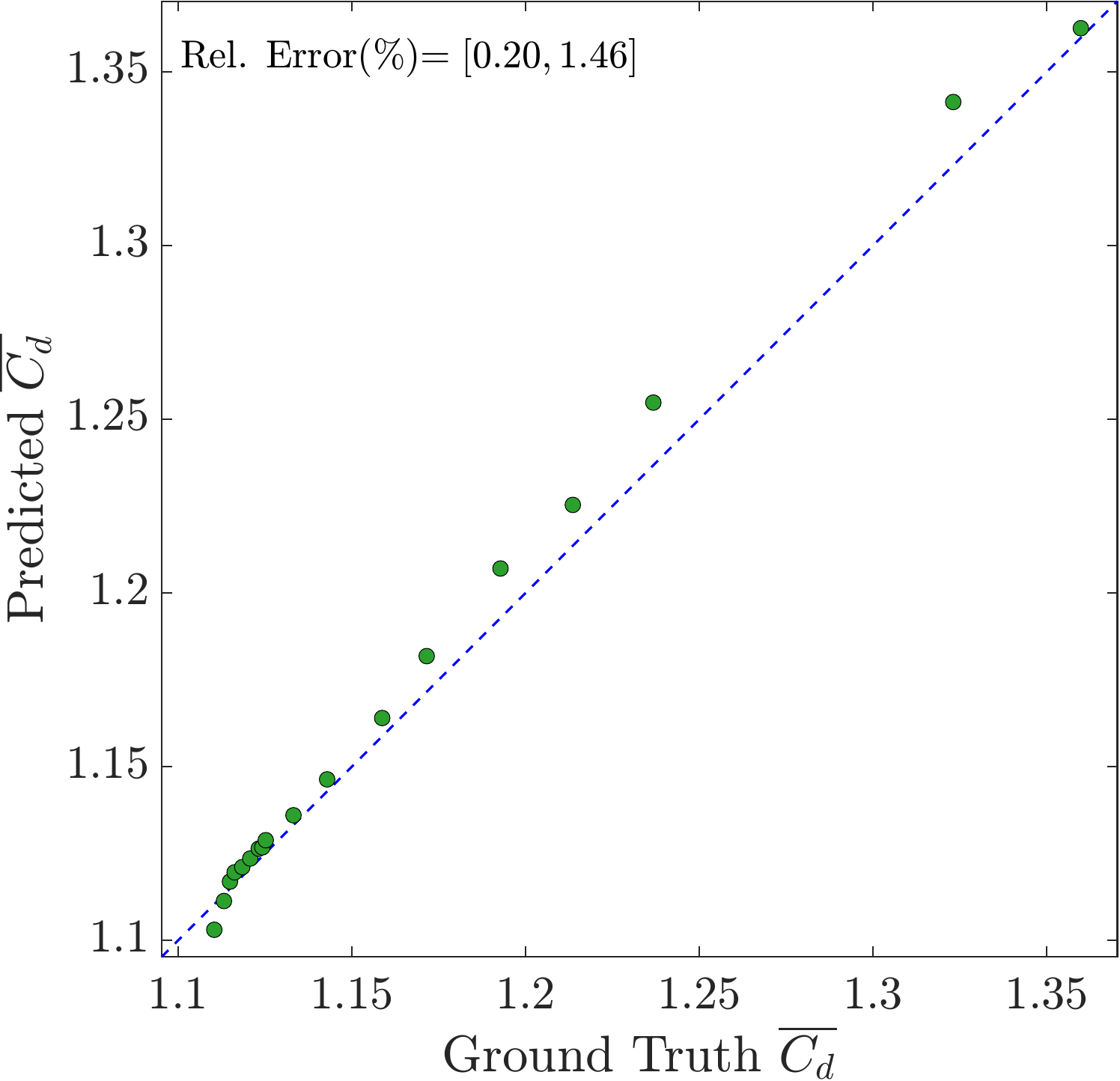}
\caption{Comparison between predicted and ground truth values for the mean coefficient of drag force on the foil}\label{fig:cdmeanc}
\end{figure}

To quantify the accuracy of the fluid states over time, we compute the coefficient of determination $R^2$ for the predicted and ground truth fields of pressure $p$ and velocity $\mathrm{[u_x, u_y]}$ across the $m^{*} = [0.015, 0.95]$  are shown in Fig. \ref{fig:R2}. The coefficient of determination is specified as:
\begin{equation}
    R^2  = 1-\frac{||\mathrm{q}-\mathrm{\hat{q}}||^{2}_{2}}{||\mathrm{q}-\mathrm{\overline{q}}||^{2}_{2}},
\end{equation}
where $\mathrm{q}$ denotes the true system state, $\mathrm{\hat{q}}$ the predicted state and $\mathrm{\overline{q}}$, the mean of the true state. Figure \ref{fig:R2} illustrates the mean $R^2$ computed here across all cases in the test dataset. The dataset is divided into three subsets based on the values of $m^*$ as listed in each sub-figure of Fig. \ref{fig:R2}. The results show that the model maintains a high predictive accuracy with $R^2> 0.94$. The low $R^2$ values for $p$ and $v$ observed in Fig. \ref{fig:R2} (a) between the timestep $580-620$ can be attributed to discrepancies in phase and amplitude errors caused during solid state predictions, as shown in Fig. \ref{fig:traj}. These discrepancies lead to a slight spatial shift in the predicted flow field contours at the foil interface and in the positioning of wake vortices, as shown in Fig. \ref{fig:flowfield3}, resulting in localized drops in the $R^2$ value.  Over longer time horizons, a gradual increase in $R^2$ is observed after the time step $1000$ for $m^2>0.65$ as shown in Fig. \ref{fig:R2}(c). This behavior of $R^2$ can be attributed to the cumulative effect of the roll-out prediction errors of the solid and fluid system states. To mitigate both localized misalignment and the decay in prediction accuracy over time, one could consider introducing ground truth data to the model at intermittent intervals, thereby anchoring the model and reinforcing stability in the predictions.  
Despite these discrepancies, the high value of $R^2$ across the full $m^*$ range demonstrates the network's ability to capture the complex fluid-structure interaction dynamics with temporal coherency and predictive consistency over a wide parameter range.   
To further evaluate the flow prediction sub-network, we compared the predicted values of the mean drag coefficient $\overline{C_{d}}$ and root mean square of the fluctuating lift coefficient $C^{rms}_{l}$ against the ground truth results across the entire test dataset. Figures \ref{fig:Clrms} and \ref{fig:cdmean} show how these quantities vary with $m^*$ for both GNN predictions and ground truth simulations, while Figs. \ref{fig:Clrmsc} and \ref{fig:cdmeanc} present the point-wise comparison along with error bounds.
%
%

As discussed in Sect.\ref{subsect: Response}, an increase in $m^*$ increases the influence of structural inertia, leading to larger flapping amplitudes. The increase in flapping amplitude correspondingly results in a decrease in lift forces and an increase in drag forces on the foil\citep{Kim2013FlappingFlag, Gurugubelli2015Self-inducedFlow} as observed in Figs. \ref{fig:Clrms} and \ref{fig:cdmean}. The flow sub-network demonstrates excellent agreement in capturing these trends when compared with the ground truth data. Achieving $R^2=0.9964$ and a maximum relative error of $1.46\%$ for $\overline{C_{d}}$ as shown in Figs. \ref{fig:cdmean} and \ref{fig:cdmeanc}. For $C^{rms}_{l}$, we observe that the model qualitatively captures the variation in $C^{rms}_{l}$ with $m^*$, but marginally overpredicts $C^{rms}_{l}$ consistently across the range $m^*$, as shown in Figs. \ref{fig:Clrms} and \ref{fig:Clrmsc}. These discrepancies lead to a lower $R^2=0.9199$ for the $C^{rms}_{l}$ as compared to $\overline{C_{d}}$.However, the absolute relative errors remain between $[0.25\%, 1.11\%]$, indicating that the fluctuating lift coefficient is predicted with good accuracy. Overall, these results illustrate that the flow sub-network can consistently predict the intricate dynamics of the flow field for the entire parameter range. 


\subsection{Energy Harvesting Potential}\label{subsec:energy}
For an undamped oscillating rigid foil, the ability of the foil to extract energy from the flow can be measured by determining the amount of fluid kinetic energy in the incoming flow that is converted into spring energy. 
The total kinetic energy in the incoming flow passing through the maximum frontal area is given as
\begin{equation}
    E_{K} = 1/2\mathrm{\rho_f}U^3|A^{peak}_y|W\hat{T},
\end{equation}
where $|A^{peak}_y|W$ represents the maximum frontal area of the foil on either the positive or negative $y$-side. $\hat{T}$ is the time taken by the foil to cross $y=0$ and reach the maximum deformation position.
The conversion ratio from fluid kinetic energy to strain energy, also known as the effectiveness of energy transfer\citep{Kim2013FlappingFlag,Gurugubelli2015EnergyFoil}, is defined as:
\begin{equation}
    R = \frac{\left(\Delta E_{s}\right)^{max}}{E_{K}} =\frac{1/2K_{\theta}\left(\theta^{peak}\right)^2}{1/2\mathrm{\rho_f}U^3|A^{peak}_y|W\hat{T}},
\end{equation}
here $\left(\Delta E_{s}\right)^{max}$ represents the maximum change in strain energy per oscillation.

\begin{figure}[htbp]
\centering\includegraphics[width=0.65\linewidth]{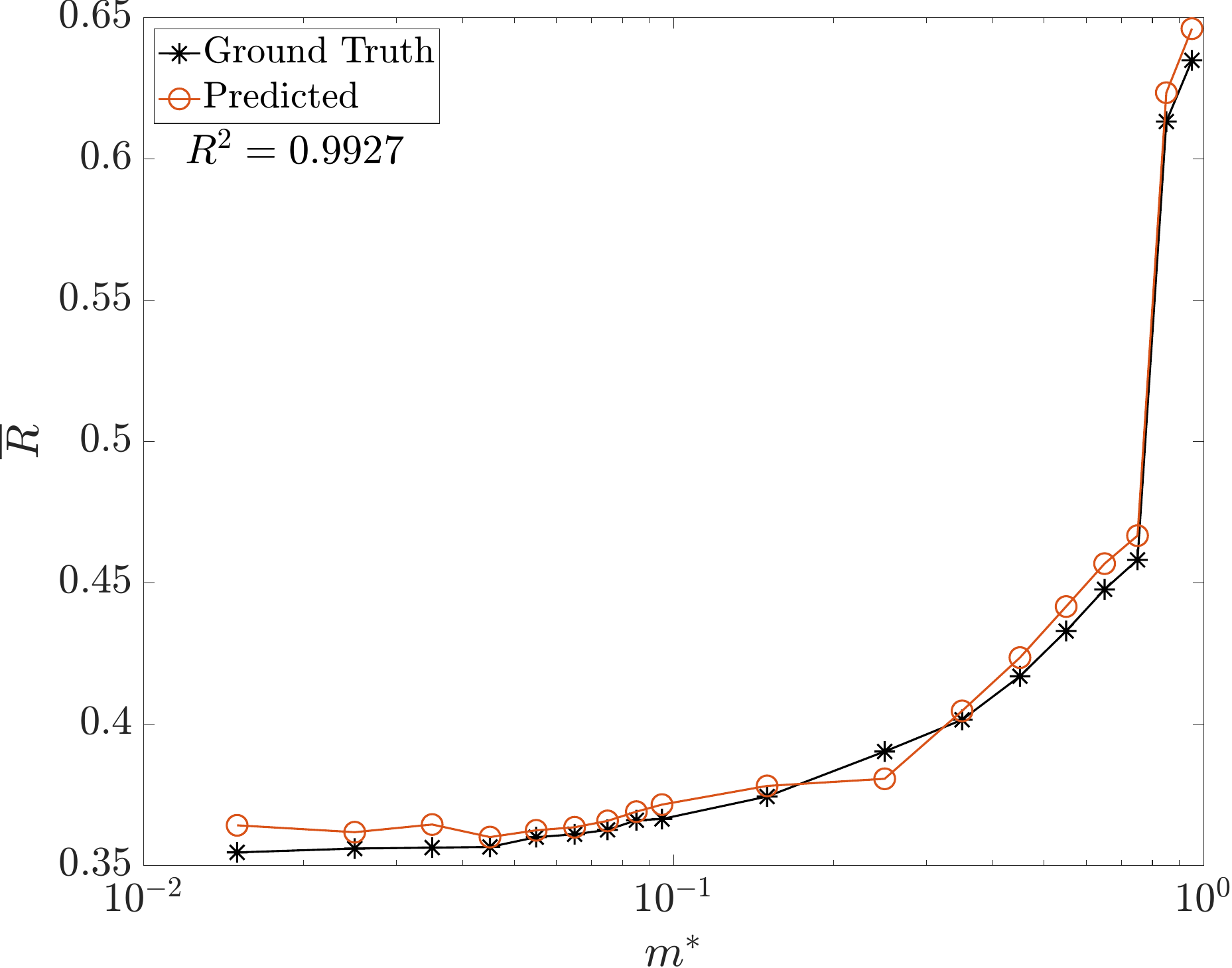}
\caption{Variation of mean effectiveness of energy transfer for the rigid foil as a function of $m^*$}\label{fig:eta}
\end{figure}

\begin{figure}[htbp]
\centering\includegraphics[width=0.55\linewidth]{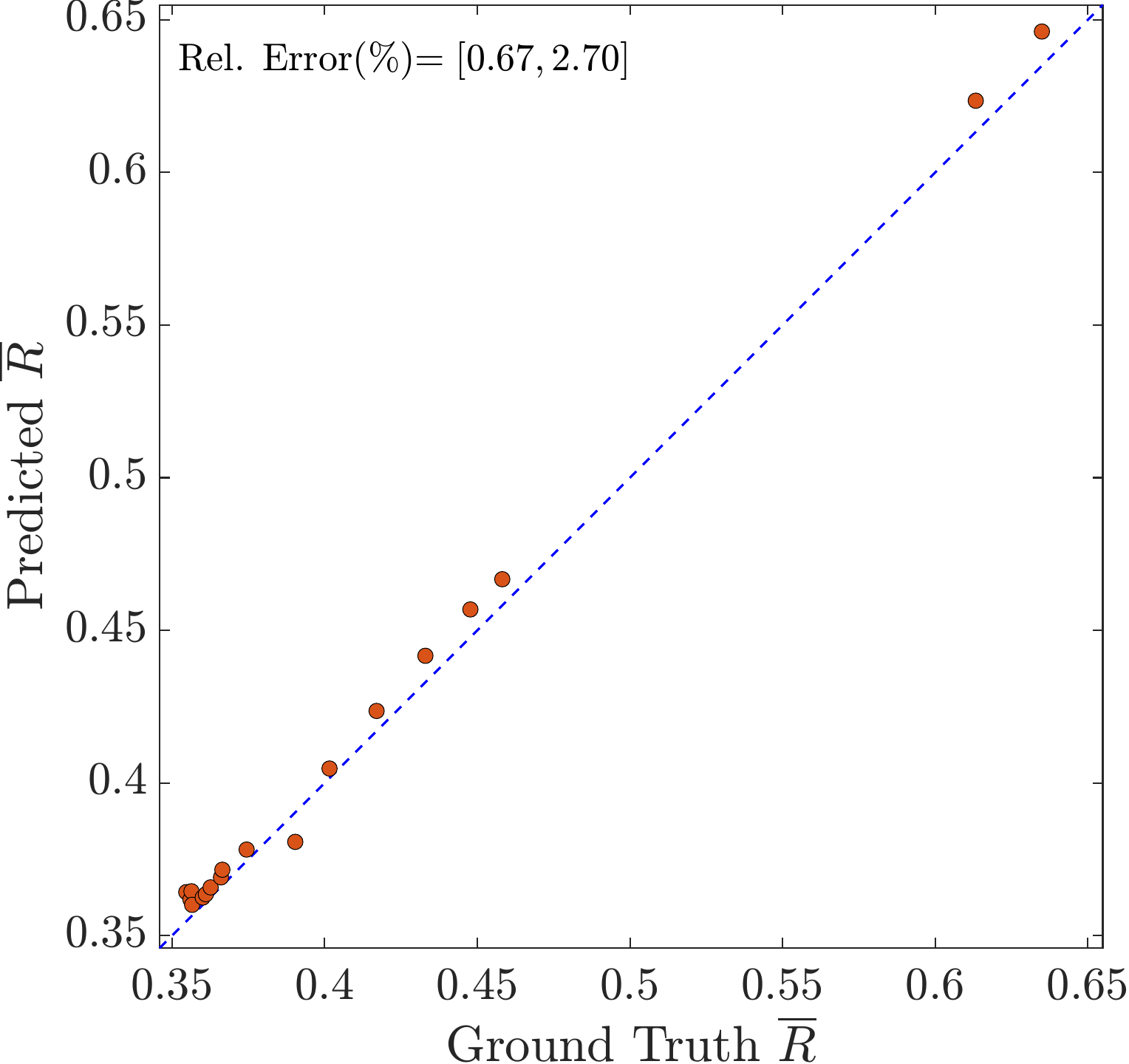}
\caption{Comparison between predicted and ground truth values for the mean effectiveness of energy transfer for the entire test set}\label{fig:etac}
\end{figure}

Figures \ref{fig:eta} and \ref{fig:etac} illustrate the model's performance in predicting the mean effectiveness of energy transfer $\overline{R}$. Here  $\overline{R}$ is computed by taking the mean effectiveness $R$ measured for each half-cycle of the foil. Figures \ref{fig:eta}
shows the variation of mean effectiveness of energy transfer $\overline{R}$ as a function of $m^*$ for both GNN predictions and ground truth, highlighting the model's ability in capturing the influence of $m^*$ on the foil's effectiveness in transferring fluid kinetic energy. Fig. \ref{fig:etac} presents a point-wise comparison between the predicted and ground truth values of $\overline{R}$, along with the absolute relative error bounds across the entire training dataset, offering a detailed overview of the prediction accuracy and consistency. As observed in these results and discussed in the earlier subsections, the increase in structural inertia with increasing $m^*$ leads to larger flapping amplitude, which in turn causes an increase in $\overline{R}$. As shown in Fig. \ref{fig:eta}, the model successfully captures the correlation between $\overline{R}$ and $m^*$, as the predictions align well with the ground truth values with an $R^2=0.9927$.  A closer look at the predicted $\overline{R}$ values reveals that the model accurately and consistently predicts effectiveness, with a maximum relative error of $2.2\%$ across the entire test dataset. In general, slightly higher relative errors in predicting effectiveness can be attributed to the cumulative effect of errors in amplitude predictions. Despite these discrepancies, the model remains effective in capturing the key trends in the amount of fluid kinetic energy stored in the spring across different mass ratios.  It is important to note that $R$ is a measure of how elastic energy is stored in the foil for possible energy transfer to other types of usable energy, and its effect on flapping dynamics should be investigated (e.g., using piezoelectric materials), which is beyond the scope of this paper.

\section{Conclusion}\label{Sect:5}
In this work, we present a deep learning-based surrogate model to predict the large-amplitude flapping response of a rigid inverted foil under uniform flow. The proposed framework employs a rotation-equivariant, quasi-monolithic hypergraph neural network to model the systems' fluid-structure interaction dynamics. Through node-element hypergraphs constructed from the unstructured computational mesh, the framework can map the higher-order relationship information intrinsic to fluid-structure interactions. The architecture is based on the arbitrary Lagrangian-Eulerian formulation, comprising of two sub-networks: a multi-layered perceptron for predicting the temporal evolution of the low-order complex-valued POD coefficients of the mesh displacements, and a hypergraph neural network for evolving the fluid flow field based on the current timestep. The surrogate model is trained and tested using high-fidelity simulation data spanning over a range of mass ratios.

Beginning from the system state at a given time step with a short temporal history of mesh states, the model produces accurate and stable roll-out predictions of the foil's response, maintaining strong agreement with ground truth data.
The model demonstrates excellent predictive performance, consistent across a range of mass ratios. Maximum relative errors of up to $1.3\%$ in transverse tip displacement, $1.1\%$ in fluid force coefficients, and $2.7\%$ in energy efficiency, highlighting its reliability for long-duration predictions. These results suggest that the surrogate model can serve as a computationally efficient alternative to high-fidelity simulations, making it valuable for real-time analysis and large-scale parametric studies.

As a part of our future work, we aim to extend this framework to more complex scenarios involving arrays of multiple inverted foils, where proximity and wake interference effects lead to nonlinear interactions that strongly affect the system's vibrational response and energy harvesting potential. Capturing these interactions will further enhance the model's applicability, enabling the fast and reliable evaluation of energy harvesting performance in practical engineering settings. Overall, by providing a scalable, data-driven approach of modeling multiphysics systems, this work takes a step towards real-time monitoring, design optimization and control of energy harvesting systems.

\section*{Acknowledgments}
The authors would like to acknowledge the Natural Sciences and Engineering Research Council of Canada (NSERC) for funding the project. The research was enabled in part through computational resources and services provided by Compute Canada and the Advanced Research Computing facility at the University of British Columbia.

\section*{Conflict of Interest}
The authors declare that they have no conflicts of interest.





\bibliographystyle{elsarticle-num} 
\bibliography{ref}

@PREAMBLE{
 "\providecommand{\noopsort}[1]{}" 
 # "\providecommand{\singleletter}[1]{#1}%" 
}

@article{allen2001energy,
  title={Energy harvesting eel},
  author={Allen, JJ and Smits, AJ},
  journal={Journal of fluids and structures},
  volume={15},
  number={3-4},
  pages={629--640},
  year={2001},
  publisher={Elsevier}
}

@article{Orrego2017HarvestingFlag,
    title = {{Harvesting ambient wind energy with an inverted piezoelectric flag}},
    year = {2017},
    journal = {Applied Energy},
    author = {Orrego, Santiago and Shoele, Kourosh and Ruas, Andre and Doran, Kyle and Caggiano, Brett and Mittal, Rajat and Kang, Sung Hoon},
    pages = {212--222},
    volume = {194},
    publisher = {Elsevier Ltd},
    issn = {03062619}
}

@article{Alam2021EnergyFoil,
    title = {{Energy harvesting from passive oscillation of inverted foil}},
    year = {2021},
    journal = {Physics of Fluids},
    author = {Alam, Md Mahbub and Chao, Li Ming and Rehman, Shafiqur and Ji, Chunning and Wang, Hanfeng},
    number = {7},
    month = {7},
    volume = {33},
    publisher = {American Institute of Physics Inc.},
    issn = {10897666}
}

@article{Goza2018GlobalFlapping,
    title = {{Global modes and nonlinear analysis of inverted-flag flapping}},
    year = {2018},
    journal = {Journal of Fluid Mechanics},
    author = {Goza, Andres and Colonius, Tim and Sader, John E.},
    month = {12},
    pages = {312--344},
    volume = {857},
    publisher = {Cambridge University Press},
    doi = {10.1017/jfm.2018.728},
    issn = {14697645},
    arxivId = {1709.09745}
}

@article{Shoele2016EnergyFlag,
    title = {{Energy harvesting by flow-induced flutter in a simple model of an inverted piezoelectric flag}},
    year = {2016},
    journal = {Journal of Fluid Mechanics},
    author = {Shoele, Kourosh and Mittal, Rajat},
    month = {3},
    pages = {582--606},
    volume = {790},
    publisher = {Cambridge University Press},
    issn = {14697645}
}

@article{Uihlein2016WaveTechnology,
    title = {{Wave and tidal current energy - A review of the current state of research beyond technology}},
    year = {2016},
    journal = {Renewable and Sustainable Energy Reviews},
    author = {Uihlein, Andreas and Magagna, Davide},
    month = {5},
    pages = {1070--1081},
    volume = {58},
    publisher = {Elsevier Ltd},
    issn = {18790690}
}

@article{bukka2021assessment,
  title={Assessment of unsteady flow predictions using hybrid deep learning based reduced-order models},
  author={Bukka, Sandeep Reddy and Gupta, Rachit and Magee, Allan Ross and Jaiman, Rajeev Kumar},
  journal={Physics of Fluids},
  volume={33},
  number={1},
  year={2021},
  publisher={AIP Publishing}
}

@article{gupta2022three,
  title={Three-dimensional deep learning-based reduced order model for unsteady flow dynamics with variable Reynolds number},
  author={Gupta, Rachit and Jaiman, Rajeev},
  journal={Physics of Fluids},
  volume={34},
  number={3},
  year={2022},
  publisher={AIP Publishing}
}

@article{mallik2022predicting,
  title={Predicting transmission loss in underwater acoustics using convolutional recurrent autoencoder network},
  author={Mallik, Wrik and Jaiman, Rajeev K and Jelovica, Jasmin},
  journal={The Journal of the Acoustical Society of America},
  volume={152},
  number={3},
  pages={1627--1638},
  year={2022},
  publisher={AIP Publishing}
}

@article{gupta2022hybrid,
  title={A hybrid partitioned deep learning methodology for moving interface and fluid--structure interaction},
  author={Gupta, Rachit and Jaiman, Rajeev},
  journal={Computers \& Fluids},
  volume={233},
  pages={105239},
  year={2022},
  publisher={Elsevier}
}

@inproceedings{miyanawala2019hybrid,
  title={A hybrid data-driven deep learning technique for fluid-structure interaction},
  author={Miyanawala, TP and Jaiman, Rajeev K},
  booktitle={International conference on offshore mechanics and arctic engineering},
  volume={58776},
  pages={V002T08A004},
  year={2019},
  organization={American Society of Mechanical Engineers}
}

@article{pfaff2020learning,
  title={Learning mesh-based simulation with graph networks},
  author={Pfaff, Tobias and Fortunato, Meire and Sanchez-Gonzalez, Alvaro and Battaglia, Peter W},
  journal={arXiv preprint arXiv:2010.03409},
  year={2020}
}

@article{gao2024finite,
  title={A finite element-inspired hypergraph neural network: Application to fluid dynamics simulations},
  author={Gao, Rui and Deo, Indu Kant and Jaiman, Rajeev K},
  journal={Journal of Computational Physics},
  volume={504},
  pages={112866},
  year={2024},
  publisher={Elsevier}
}

@article{xu2021conditionally,
  title={Conditionally parameterized, discretization-aware neural networks for mesh-based modeling of physical systems},
  author={Xu, Jiayang and Pradhan, Aniruddhe and Duraisamy, Karthikeyan},
  journal={Advances in Neural Information Processing Systems},
  volume={34},
  pages={1634--1645},
  year={2021}
}

@article{gao2024towards,
  title={Towards spatio-temporal prediction of cavitating fluid flow with graph neural networks},
  author={Gao, Rui and Heydari, Shayan and Jaiman, Rajeev K},
  journal={International Journal of Multiphase Flow},
  volume={177},
  pages={104858},
  year={2024},
  publisher={Elsevier}
}

@inproceedings{reddy2019reduced,
  title={Reduced order model for unsteady fluid flows via recurrent neural networks},
  author={Reddy, Sandeep B and Magee, Allan Ross and Jaiman, Rajeev K and Liu, J and Xu, W and Choudhary, A and Hussain, AA},
  booktitle={International Conference on Offshore Mechanics and Arctic Engineering},
  volume={58776},
  pages={V002T08A007},
  year={2019},
  organization={American Society of Mechanical Engineers}
}

@article{chen2023towards,
  title={Towards high-accuracy deep learning inference of compressible flows over aerofoils},
  author={Chen, Li-Wei and Thuerey, Nils},
  journal={Computers \& Fluids},
  volume={250},
  pages={105707},
  year={2023},
  publisher={Elsevier}
}

@article{jaiman2016stable,
  title={A stable second-order partitioned iterative scheme for freely vibrating low-mass bluff bodies in a uniform flow},
  author={Jaiman, RK and Pillalamarri, NR and Guan, MZ},
  journal={Computer Methods in Applied Mechanics and Engineering},
  volume={301},
  pages={187--215},
  year={2016},
  publisher={Elsevier}
}

@article{jaiman2016partitioned,
  title={Partitioned iterative and dynamic subgrid-scale methods for freely vibrating square-section structures at subcritical Reynolds number},
  author={Jaiman, Rajeev K and Guan, MZ and Miyanawala, TP},
  journal={Computers \& Fluids},
  volume={133},
  pages={68--89},
  year={2016},
  publisher={Elsevier}
}

@article{Liu2014AEffects,
    title = {{A stable second-order scheme for fluid-structure interaction with strong added-mass effects}},
    year = {2014},
    journal = {Journal of Computational Physics},
    author = {Liu, Jie and Jaiman, Rajeev K. and Gurugubelli, Pardha S.},
    month = {8},
    pages = {687--710},
    volume = {270},
    publisher = {Academic Press Inc.},
    issn = {10902716}
}

@article{gao2024predicting,
  title={Predicting fluid--structure interaction with graph neural networks},
  author={Gao, Rui and Jaiman, Rajeev K},
  journal={Physics of Fluids},
  volume={36},
  number={1},
  year={2024},
  publisher={AIP Publishing}
}

@article{Kim2013FlappingFlag,
    title = {{Flapping dynamics of an inverted flag}},
    year = {2013},
    journal = {Journal of Fluid Mechanics},
    author = {Kim, Daegyoum and Coss{\'{e}}, Julia and Cerdeira, Cecilia and Gharib, Morteza},
    pages = {10},
    publisher = {Cambridge University Press}
}

@article{Gurugubelli2015Self-inducedFlow,
    title = {{Self-induced flapping dynamics of a flexible inverted foil in a uniform flow}},
    year = {2015},
    journal = {Journal of Fluid Mechanics},
    author = {Gurugubelli, P. S. and Jaiman, R. K.},
    month = {10},
    pages = {657--694},
    volume = {781},
    publisher = {Cambridge University Press},
    issn = {14697645}
}

@article{leontini2022dynamics,
  title={The dynamics of a rigid inverted flag},
  author={Leontini, Justin S and Sader, John E},
  journal={Journal of Fluid Mechanics},
  volume={948},
  pages={A47},
  year={2022},
  publisher={Cambridge University Press}
}

@article{Gurugubelli2019LargePeriodicity,
    title = {{Large amplitude flapping of an inverted elastic foil in uniform flow with spanwise periodicity}},
    year = {2019},
    journal = {Journal of Fluids and Structures},
    author = {Gurugubelli, P. S. and Jaiman, R. K.},
    month = {10},
    pages = {139--163},
    volume = {90},
    publisher = {Academic Press},
    doi = {10.1016/j.jfluidstructs.2019.05.009},
    issn = {10958622}
}

@inproceedings{Gurugubelli2015EnergyFoil,
    title = {{Energy Harvesting using flapping dynamics of piezoelectric inverted flexible foil}},
    year = {2015},
    booktitle = {International Conference on Offshore Mechanics and Arctic Engineering},
    author = {Gurugubelli, Pardha S and Jaiman, Rajeev K},
    pages = {V009T09A010},
    volume = {56574},
    publisher = {American Society of Mechanical Engineers},
    organization = {American Society of Mechanical Engineers}
}

@article{virtanen2020scipy,
  title={SciPy 1.0: fundamental algorithms for scientific computing in Python},
  author={Virtanen, Pauli and Gommers, Ralf and Oliphant, Travis E and Haberland, Matt and Reddy, Tyler and Cournapeau, David and Burovski, Evgeni and Peterson, Pearu and Weckesser, Warren and Bright, Jonathan and others},
  journal={Nature methods},
  volume={17},
  number={3},
  pages={261--272},
  year={2020},
  publisher={Nature Publishing Group US New York}
}

@article{sitzmann2020implicit,
  title={Implicit neural representations with periodic activation functions},
  author={Sitzmann, Vincent and Martel, Julien and Bergman, Alexander and Lindell, David and Wetzstein, Gordon},
  journal={Advances in neural information processing systems},
  volume={33},
  pages={7462--7473},
  year={2020}
}

@article{paszke2019pytorch,
  title={Pytorch: An imperative style, high-performance deep learning library},
  author={Paszke, Adam and Gross, Sam and Massa, Francisco and Lerer, Adam and Bradbury, James and Chanan, Gregory and Killeen, Trevor and Lin, Zeming and Gimelshein, Natalia and Antiga, Luca and others},
  journal={Advances in neural information processing systems},
  volume={32},
  year={2019}
}

@inproceedings{pascanu2013difficulty,
  title={On the difficulty of training recurrent neural networks},
  author={Pascanu, Razvan and Mikolov, Tomas and Bengio, Yoshua},
  booktitle={International conference on machine learning},
  pages={1310--1318},
  year={2013},
  organization={Pmlr}
}

@article{kingma2014adam,
  title={Adam: A method for stochastic optimization},
  author={Kingma, Diederik P and Ba, Jimmy},
  journal={arXiv preprint arXiv:1412.6980},
  year={2014}
}

@article{tavallaeinejad2021dynamics,
  title={Dynamics of inverted flags: Experiments and comparison with theory},
  author={Tavallaeinejad, Mohammad and Salinas, Manuel Flores and Pa{\"\i}doussis, Michael P and Legrand, Mathias and Kheiri, Mojtaba and Botez, Ruxandra M},
  journal={Journal of Fluids and Structures},
  volume={101},
  pages={103199},
  year={2021},
  publisher={Elsevier}
}

@article{parekh2025wake,
  title={Wake interference effects on flapping dynamics of elastic inverted foil},
  author={Parekh, Aarshana R and Jaiman, Rajeev K},
  journal={Physical Review Fluids},
  volume={10},
  number={1},
  pages={014702},
  year={2025},
  publisher={APS}
}

@article{pant2021deep,
  title={Deep learning for reduced order modelling and efficient temporal evolution of fluid simulations},
  author={Pant, Pranshu and Doshi, Ruchit and Bahl, Pranav and Barati Farimani, Amir},
  journal={Physics of Fluids},
  volume={33},
  number={10},
  year={2021},
  publisher={AIP Publishing}
}

@article{thuerey2020deep,
  title={Deep learning methods for Reynolds-averaged Navier--Stokes simulations of airfoil flows},
  author={Thuerey, Nils and Wei{\ss}enow, Konstantin and Prantl, Lukas and Hu, Xiangyu},
  journal={AIAA journal},
  volume={58},
  number={1},
  pages={25--36},
  year={2020},
  publisher={American Institute of Aeronautics and Astronautics}
}

@article{hadizadeh2024graph,
  title={A graph neural network surrogate model for multi-objective fluid-acoustic shape optimization},
  author={Hadizadeh, Farnoosh and Mallik, Wrik and Jaiman, Rajeev K},
  journal={arXiv preprint arXiv:2412.16817},
  year={2024}
}

@article{lino2022multi,
  title={Multi-scale rotation-equivariant graph neural networks for unsteady Eulerian fluid dynamics},
  author={Lino, Mario and Fotiadis, Stathi and Bharath, Anil A and Cantwell, Chris D},
  journal={Physics of Fluids},
  volume={34},
  number={8},
  year={2022},
  publisher={AIP Publishing}
}

@inproceedings{belbute2020combining,
  title={Combining differentiable PDE solvers and graph neural networks for fluid flow prediction},
  author={Belbute-Peres, Filipe De Avila and Economon, Thomas and Kolter, Zico},
  booktitle={international conference on machine learning},
  pages={2402--2411},
  year={2020},
  organization={PMLR}
}

@article{Watanabe2002AFlutter,
    title = {{A theoretical study of paper flutter}},
    year = {2002},
    journal = {Journal of Fluids and Structures},
    author = {Watanabe, Y. and Isogai, K. and Suzuki, S. and Sugihara, M.},
    number = {4},
    pages = {543--560},
    volume = {16},
    doi = {10.1006/jfls.2001.0436},
    issn = {08899746}
}

@techreport{Guo2000StabilityFlow,
    title = {{Stability of Rectangular Plates With Free Side-Edges in Two-Dimensional Inviscid Channel Flow}},
    year = {2000},
    author = {Guo, C Q and Paidoussis, M P and Asme, Fellow},
}

@article{lucey1998excitation,
  title={The excitation of waves on a flexible panel in a uniform flow},
  author={Lucey, Anthony D},
  journal={Philosophical Transactions of the Royal Society of London. Series A: Mathematical, Physical and Engineering Sciences},
  volume={356},
  number={1749},
  pages={2999--3039},
  year={1998},
  publisher={The Royal Society}
}

@article{Michelin2013EnergyFlows,
    title = {{Energy harvesting efficiency of piezoelectric flags in axial flows}},
    year = {2013},
    journal = {Journal of Fluid Mechanics},
    author = {Michelin, S. and Doar{\'{e}}, D.},
    month = {1},
    pages = {489--504},
    volume = {714},
    publisher = {Cambridge University Press},
    doi = {10.1017/jfm.2012.494},
    issn = {14697645},
    arxivId = {1210.2171}
}

@article{deng2022influence,
  title={Influence of tandem fluttering membranes on flow dynamics and heat transfer in turbulent channel flow},
  author={Deng, Yifan and Tang, Yuchao and Wang, Peng and Cao, Zhaomin and Liu, Yingzheng},
  journal={Physics of Fluids},
  volume={34},
  number={1},
  year={2022},
  publisher={AIP Publishing}
}

@article{ali2015heat,
  title={Heat transfer and mixing enhancement by free elastic flaps oscillation},
  author={Ali, Samer and Habchi, Charbel and Menanteau, S{\'e}bastien and Lemenand, Thierry and Harion, Jean-Luc},
  journal={International Journal of Heat and Mass Transfer},
  volume={85},
  pages={250--264},
  year={2015},
  publisher={Elsevier}
}

@article{Zhang2000FlexibleWind,
    title = {{Flexible filaments in a flowing soap film as a model for one-dimensional flags in a two-dimensional wind}},
    year = {2000},
    journal = {Nature},
    author = {Zhang, Jun and Childress, Stephen and Libchaber, Albert and Shelley, Michael},
    number = {6814},
    pages = {835--839},
    volume = {408}
}







\end{document}